\documentclass[12pt,dvips]{article}

\usepackage{rotating}
\usepackage{axodraw}
\usepackage{epsfig}
\usepackage{epsf}
\usepackage{psfig}
\usepackage{cite}

\newcommand{\imag}{\Im {\rm m}}
\newcommand{\real}{\Re {\rm e}}

\begin{document}

\begin{flushright}
CERN-PH-TH/2004-051 \\
MC-TH-2004-03 \\
{\tt hep-ph/0404167} \\
April 2004
\end{flushright}

\begin{center}
{\bf {\LARGE LHC Signatures of Resonant CP Violation }}\\[3mm] {\bf
{\LARGE in a Minimal Supersymmetric Higgs Sector}}
\end{center}

\medskip

\begin{center}
{\large John Ellis$^{\,a}$, Jae Sik Lee$^{\,b}$ 
                                       and Apostolos Pilaftsis$^{\,b}$}
\end{center}

\begin{center}
{\em $^a$Theory Division, Physics Department, CERN, CH-1211 Geneva 23, 
Switzerland}\\[2mm]
{\em $^b$Department of Physics and Astronomy, University of Manchester}\\
{\em Manchester M13 9PL, United Kingdom}
\end{center}

\bigskip 
\bigskip 
\bigskip 

\centerline{\bf ABSTRACT}
\medskip
\noindent  
We present  the general formalism for  studying CP-violating phenomena
in  the  production,   mixing  and  decay  of  a   coupled  system  of
CP-violating   neutral   Higgs   bosons  at   high-energy   colliders.
Considering  the Minimal  Supersymmetric Standard  Model  (MSSM) Higgs
sector in which  CP violation is radiatively induced  by phases in the
soft   supersymmetry-breaking    third-generation   trilinear   squark
couplings and gaugino masses, we  apply our formalism to neutral Higgs
production via ${\bar b}b$, $gg$  and $W^+ W^-$ collisions at the LHC.
We  discuss   CP  asymmetries  in  the   longitudinal  and  transverse
polarizations  of  $\tau^+  \tau^-$   pairs.   The  signatures  of  CP
violation are more prominent in  the production via $gg$ and $W^+ W^-$
than via  ${\bar b}b$,  and are resonantly  enhanced when two  (or all
three)  neutral   Higgs  bosons   are  nearly  degenerate   with  mass
differences comparable  to their  decay widths.  Such  scenarios occur
naturally in the MSSM  for values of $\tan \beta \stackrel{>}{{}_\sim}
5~(30)$  and large  (small)  charged Higgs-boson  masses.  We  analyze
representative examples  with large  mixing between the  three neutral
Higgs bosons  weighing about 120~GeV,  that may exhibit  observable CP
asymmetries even as large as~80\%.

\newpage

\setcounter{equation}{0}
\section{Introduction}
\label{sec:introduction}

If supersymmetry  (SUSY) turns  out to  be  realized  at low  energies
$\stackrel{<}{{}_\sim}  1$~TeV~\cite{HPN}, the following   interesting
questions will then arise: does SUSY  make observable contributions to
the  violation   of either  flavour  or  CP?     Even  in  the minimal
supersymmetric extension   of the  Standard   Model (MSSM),  the  soft
SUSY-breaking sector may include    about a hundred   parameters  that
violate these symmetries. However, if one imposes flavour conservation
on  the soft  SUSY-breaking parameters  $m_0,   m_{1/2}$ and  $A$, and
assumes  that they are universal,  then only two physical CP-violating
phases  remain: one  in the  gaugino  masses $m_{1/2}$ and  one in the
trilinear couplings $A$.

These CP-violating phases may in principle be measured directly in the
production  cross   sections  and   decay  widths  of   sparticles  at
high-energy colliders~\cite{CPdirect,CPsoft},  or indirectly via their
radiative effect  on the Higgs sector~\cite{APLB}~\footnote{Additional
indirect  constraints on the  soft SUSY-breaking  phases and  the MSSM
mass  spectrum may be  obtained from  experimental limits  on electric
dipole     moments    (EDMs)~\cite{EDM1,EDM2,CKP}     and    $B$-meson
observables~\cite{Bmeson1,DP}.}.   The  Higgs sector  of  the MSSM  is
affected    at     the    one-loop    level     by    the    trilinear
phase~\cite{APLB,PW,Demir,CDL,CEPW,INhiggs,KW,HeinCP,CEPW2} and at the
two-loop        level         by        the        gaugino        mass
phase~\cite{CDL,CEPW,INhiggs,CEPW2}.   This loop-induced  CP violation
mixes the CP-even Higgses $h,H$ with the CP-odd Higgs boson $A$.  Many
studies have  been made of the  masses and couplings  of the resulting
mixed-CP Higgs bosons $H_{1,2,3}$,  and some of their phenomenological
consequences for searches  at LEP and future colliders  have also been
considered~\cite{CHL,CPX,CEMPW,CPpp,CFLMP,KMR,CPee,CPphoton,CPmumu}.

More  complete studies  of CP-violating  Higgs bosons  will  require a
careful treatment of the resonant mixing of multiple Higgs bosons that
couple to  the same initial and  final states.  In  general, one could
expect that the  CP-violating mixing of the heavier  MSSM Higgs bosons
$H,  A$ may be  more important  than their  mixings with  the lightest
CP-even  Higgs boson  $h$.  However, non-negligible  mixing among  all
three neutral Higgs states is  also possible in a general CP-violating
MSSM.  Such a scenario naturally  emerges from a parameter space where
$\tan\beta$ is  large, i.e.~$\tan\beta \stackrel{>}{{}_\sim}  30$, and
the charged Higgs bosons  $H^\pm$ are relatively light with $M_{H^\pm}
\stackrel{<}{{}_\sim} 160$~GeV.

In this  paper, we  develop the general  formalism for  describing the
dynamics that  governs the production,  mixing and decay of  a coupled
system of CP-violating neutral  Higgs bosons.  Our formalism makes use
of the field-theoretic  resummation approach developed in~\cite{APNPB}
to treat unstable particle-mixing  transitions.  Within the context of
gauge  theories,  it  is  important  that  resummation  approaches  to
unstable  particles   consistently  maintain  crucial  field-theoretic
properties,    such    as    gauge   invariance,    analyticity    and
unitarity~\cite{PP}.  It has been shown in~\cite{PP,BP} that all these
properties   are  preserved   within  the   framework  of   the  Pinch
Technique~(PT)~\cite{PTgeneral}.  Here, using the  PT, we  compute the
gauge-mediated diagonal  as well  as off-diagonal absorptive  parts in
the  resummed  Higgs-boson propagator  matrix.  Finally, an  essential
ingredient of our formalism is  the inclusion of the CP-violating loop
corrections in the production and decay vertices of the Higgs bosons.

We   illustrate  our   general  formalism   for   the  coupled-channel
$H_{1,2,3}$ mixing by explicit  treatments of the production processes
$g  g$, $b  {\bar  b}$~\footnote{We note  that  the $b\bar{b}$  fusion
process may become the leading production channel at large $\tan\beta$
at the LHC,  as has recently been shown  in~\cite{BLS}.}  and $W^+ W^-
\to  H_{1,2,3}  \to \tau^+  \tau^-$.   These  are  the most  important
production mechanisms for  neutral Higgs bosons at the  LHC, while the
decay  channel  that  seems  the  most promising  for  studies  of  CP
violation is that into $\tau^+ \tau^-$ pairs.  To quantify the genuine
signatures  of CP  violation,  we calculate  CP  asymmetries that  are
defined in  terms of longitudinal and transverse  polarizations of the
$\tau^\pm$  leptons.  When  $\tan\beta$  is large  and/or the  charged
Higgs boson mass is large, so that two or more Higgs bosons are nearly
degenerate,  even  small  CP-violating  phases could  induce  sizeable
CP-violating mixing.  However, as we demonstrate, there are systematic
cancellations  due  to  CPT-preserving  rescattering  effects  in  the
process $b {\bar b} \to H_{1,2,3} \to \tau^+ \tau^-$ that suppress the
CP-violating signatures in this case.  There are no such cancellations
in $g  g$ and $W^+ W^-  \to H_{1,2,3} \to \tau^+  \tau^-$, which could
have much larger CP asymmetries at the LHC.  We analyze representative
examples with large three-way mixing to show that these CP asymmetries
might well  exceed the 10\%  level and could  even reach values  up to
80\%.

Although our  predictions are  obtained in the  MSSM with  explicit CP
violation, it  is important to stress that  large CP-violating effects
could also occur in  a general CP-violating 2-Higgs-doublet model with
a similar Higgs-boson mass  spectrum. Our presentation is organized in
such  a  way  that the  formalism  may  easily  be extended  to  Higgs
production  at other colliders.   For instance,  our formalism  can be
applied to  $\gamma \gamma$ colliders~\cite{CPphoton,TauFusion}, which
are  analogous  to $g  g$  collisions at  the  LHC,  to $\mu^+  \mu^-$
colliders~\cite{CPmumu}, which have  formal similarities with $b {\bar
b}$  collisions at  the  LHC, and  to  $WW$-fusion and  Higgsstrahlung
processes at $e^+e^-$ linear colliders~\cite{Marek}.

Section~\ref{sec:formalism}  presents
the general formalism for     the coupled-channel analysis   of  Higgs
bosons, including  explicit formulae for  the  absorptive parts of the
Higgs-boson  propagator  matrix   and   the  vertex corrections.    In
Section~\ref{sec:production}, we apply the results of our formalism to
the production  channels $gg,b\bar{b},W^+W^-  \to \tau^+\tau^-$ at the
LHC.  In Section~\ref{sec:numerical} we present numerical estimates of
particular CP-violating   MSSM   scenarios  that  exhibit  large    CP
asymmetries.  Our numerical estimates are   based on the Fortran  code
{\tt                CPsuperH}~\cite{CPsuperH}.                Finally,
Section~\ref{sec:conclusions} contains   our conclusions and discusses
the  prospects for pursuing studies  of  Higgs-sector CP violation  at
future colliders beyond the~LHC.

\setcounter{equation}{0}
\section{Formalism for Coupled-Channel Analyses of Higgs-Sector CP 
Violation}
\label{sec:formalism}  

We consider situations where two  or more MSSM Higgs bosons contribute
simultaneously to  the  production  of some  fermion-antifermion  pair
whose polarization states can  be measured. We treat explicitly
the example of $H_{1,2,3} \to \tau^+  \tau^-$, but the formalism could
easily   be adapted to   other cases  such  as  $t {\bar t},  \chi^+_i
\chi^-_j$  or    $\chi^0_i \chi^0_j$.    There   have   been extensive
discussions of the masses and couplings  of MSSM Higgs bosons mixed by
loop-induced     CP violation~\cite{PW,Demir,CDL,CEPW}.  To   account 
properly for the constraints  that CPT invariance and unitarity imposes
on the   cross   sections~\cite{APNPB},  we must   consider   the full
off-shell propagator matrix  for mixed  MSSM  Higgs bosons,  including
off-diagonal absorptive parts.

The  absorptive part  of  the Higgs-boson  propagator matrix  receives
contributions from loops of  fermions, vector bosons, associated pairs
of Higgs and vector bosons, Higgs-boson pairs, and sfermions:
\begin{equation}
\imag\widehat{\Pi}_{ij}(s)= \imag\widehat{\Pi}^{ff}_{ij}(s)+
\imag\widehat{\Pi}^{VV}_{ij}(s)+\imag\widehat{\Pi}^{HV}_{ij}(s) +
\imag\widehat{\Pi}^{HH}_{ij}(s)
+ \imag\widehat{\Pi}^{\tilde{f}\tilde{f}}_{ij}(s)\,.
\end{equation}
The  contributions  of the  exchanges of   the bottom and  top quarks,
$\tau$ leptons,   neutralinos $\chi^0_i$ and  charginos $\chi^+_i$ are
summed  in $\imag\widehat{\Pi}^{ff}_{ij}(s)$.  The latter may conveniently
be cast into the form
\begin{eqnarray}
\imag\widehat{\Pi}^{ff}_{ij}(s)&=&\frac{s}{8\pi}
\sum_{f,f^\prime=b,t,\tau,\tilde{\chi}^0,\tilde{\chi}^-}
K_f(s)\, g_f^2 \Delta_{ff^\prime}N_C^f
\left[ (1-\kappa_f-\kappa_{f^\prime})
(g^S_{H_i \bar{f^\prime}f} g^{S*}_{H_j \bar{f^\prime}f}
+g^P_{H_i \bar{f^\prime}f} g^{P*}_{H_j \bar{f^\prime}f}) \right.
\nonumber \\
&& \hspace{-2cm}
\left.
-2 \sqrt{\kappa_f\kappa_{f^\prime}}
(g^S_{H_i \bar{f^\prime}f} g^{S*}_{H_j \bar{f^\prime}f}
-g^P_{H_i \bar{f^\prime}f} g^{P*}_{H_j \bar{f^\prime}f})\right]
\lambda^{1/2}(1,\kappa_f,\kappa_{f^\prime})\:
\Theta\left(s-(m_f+m_{f^\prime})^2\right),
\end{eqnarray}
where  $K_{b,t}(s)\simeq    1         +  5.67\frac{\alpha_s(s)}{\pi}$,
$\Delta_{ff^\prime}=\delta_{ff^\prime}\,(f,f^\prime=b,t,\tau)$,
$\frac{4}{1+\delta_{ff^\prime}}\,(f,f'=\tilde{\chi}^0_{1,2,3,4})$,  or
$1\,(f,f'=\tilde{\chi}^-_{1,2})$,                                  and
$\lambda(x,y,z)=x^2+y^2+z^2-2(xy+yz+zx)$       with    $\kappa_x\equiv
m_x^2/s$.  Here and  subsequently,  we follow the convention   of {\tt
CPsuperH}  \cite{CPsuperH} for the couplings   of the Higgs bosons  to
fermions, vector bosons, Higgs bosons, and sfermions.
For the calculation of the bottom- and top-quark contributions, 
the running quark
masses at the scale $\sqrt{s}$ have been used in the couplings
$g_{b,t}=gm_{b,t}(\sqrt{s})/2M_W$. Specifically, we use $m_b (m_t^{\rm
pole})=3$ GeV, where $m_t^{\rm pole}=175$ GeV.
 
The vector-boson loop contributions are
\begin{eqnarray}
  \label{hatVV}
\imag\widehat{\Pi}^{VV}_{ij}(s)&=&\frac{g^2g_{_{H_iVV}}g_{_{H_jVV}}\delta_V}
{128\pi M_W^2}
\beta_V\left[-4M_V^2(2s-3M_V^2)\right.
\nonumber \\
&& \hspace{2cm}
\left.
+2M_V^2(M_{H_i}^2+M_{H_j}^2)
+M_{H_i}^2M_{H_j}^2 \right]\Theta(s-4 M_V^2),
\end{eqnarray}
where $\beta_V=(1-4\kappa_V)^{1/2}$ and $\delta_W=2$, $\delta_Z=1$. 

Correspondingly, the exchanges of Higgs and vector boson pairs give
\begin{eqnarray}
  \label{hatHV}
\imag\widehat{\Pi}^{HV}_{ij}(s)
&=&\frac{g^2}{64\pi M_W^2}\sum_{k=1,2,3}
g_{_{H_iH_kZ}}g_{_{H_jH_kZ}}
\lambda^{1/2}(1,\kappa_Z,\kappa_{H_k})
\left[-4sM_Z^2+(M_Z^2-M_{H_k}^2)^2 \right.
\nonumber \\
&&
\left.
+(M_Z^2-M_{H_k}^2)(M_{H_i}^2+M_{H_j}^2)
+M_{H_i}^2M_{H_j}^2 \right]
\Theta\left(s- (M_Z+M_{H_k})^2\right)
\nonumber \\
&+&\frac{g^2}{32\pi M_W^2}
\real(g_{_{H_iH^+W^-}}g^*_{_{H_jH^+W^-}})
\lambda^{1/2}(1,\kappa_W,\kappa_{H^\pm})
\left[-4sM_W^2+(M_W^2-M_{H^\pm}^2)^2 \right.
\nonumber \\
&&
\left.
+(M_W^2-M_{H^\pm}^2)(M_{H_i}^2+M_{H_j}^2)
+M_{H_i}^2M_{H_j}^2 \right]
\Theta\left(s- (M_W+M_{H^\pm})^2\right)\,.
\end{eqnarray}
In  deriving (\ref{hatVV}) and (\ref{hatHV}), we   apply the PT to the
MSSM Higgs  sector following  a procedure very  analogous  to  the one
given in~\cite{PINCH} for the SM Higgs sector.   As a consequence, the
PT        self-energies   $\imag\widehat{\Pi}^{VV}_{ij}(s)$        and
$\imag\widehat{\Pi}^{VH}_{ij}(s)$   depend  linearly  on $s$   at high
energies.  This differs  crucially from the bad high-energy dependence
$ \propto  s^2$   that one usually  encounters   when  the Higgs-boson
self-energies are calculated  in the unitary gauge.   In fact, if  the
Higgs-boson self-energies  are  embedded in  a truly gauge-independent
quantity such as the S-matrix element of a $2\to 2$ process, the badly
high-energy-behaved $s^2$-dependent terms cancel against corresponding
$s^2$  terms  present in  the vertices  and boxes  order  by  order in
perturbation  theory.  In this context,  PT provides a self-consistent
approach  to extract those    $s^2$-dependent  terms from   boxes  and
vertices, thus giving rise to  effective Higgs self-energies that  are
independent of the gauge-fixing parameter  and $s^2$.  More details on
the PT may be found in~\cite{PP,BP,PTgeneral,PINCH}.

Finally, the contributions of the MSSM Higgs bosons and sfermions are
\begin{equation}
  \label{A5}
\imag\widehat{\Pi}^{HH}_{ij}(s)=
\frac{v^2}{16\pi}\sum_{k\geq l=1,2,3}
\frac{S_{ij;kl}}{1+\delta_{kl}}
g_{_{H_iH_kH_l}} g_{_{H_jH_kH_l}}
\lambda^{1/2}(1,\kappa_{H_k},\kappa_{H_l})\:
\Theta\left(s- (M_{H_k}+M_{H_l})^2\right)\,,
\end{equation}
\begin{equation}
  \label{A6}
\imag\widehat{\Pi}^{\tilde{f}\tilde{f}}_{ij}(s)=
\frac{v^2}{16\pi}\sum_{f=b,t,\tau}\sum_{k,l=1,2} N_C^f\,
g_{H_i\tilde{f}^*_k\tilde{f}_l}
g^*_{H_j\tilde{f}^*_k\tilde{f}_l}
\lambda^{1/2}(1,\kappa_{\tilde{f}_k},\kappa_{\tilde{f}_l})\:
\Theta\left(s- (M_{\tilde{f}_k}+M_{\tilde{f}_l})^2\right)\, .
\end{equation}
Note  that  the  symmetry  factor  $S_{ij;kl}$ has  to  be  calculated
appropriately.  When  $i=j=1$ and  $k=l=2$, for example,  the symmetry
factor   for  the   squared   self-coupling  $g^2_{_{H_1H_2H_2}}$   is
$S_{11;22}=4$.

When considering any   specific production process  and decay channel,
the   Higgs-boson propagator    matrix  must   be  combined  with  the
appropriate vertices,   that   themselves receive   CP-violating  loop
corrections. Since the main decay  channel we consider  for the LHC is
$\tau^+ \tau^-$, and since   many of the interesting Higgs  production
and other decay mechanisms also involve fermions such as $b {\bar b}$,
we also summarize relevant  aspects of the loop-induced corrections to
the $H_{1,2,3}f \bar{f}$ vertices.

The exchanges    of  gluinos and charginos  give   finite loop-induced
corrections to   the $H_{1,2,3}b {\bar   b}$ Yukawa coupling  with the
structure
\begin{equation}
h_b\ =\ \frac{\sqrt{2}m_b}{v\cos\beta}\,
\frac{1}{1+(\delta h_b/h_b)+(\Delta h_b/h_b)\tan\beta} \,.
\end{equation}
The $\tan\beta$-enhanced  threshold correction  $(\Delta h_b/h_b)$ has
terms proportional to the strong coupling $\alpha_s$ and the top-quark
Yukawa    coupling    $|h_t|^2$.
See Eqs.~(2.4) and (2.5) in~\cite{CEMPW} for
the analytic forms of $(\delta h_b/h_b)$ and 
$(\Delta h_b/h_b)$, respectively.
In addition,   there are contributions to   $(\Delta
h_b/h_b)$ coming  from  the exchanges of  binos  and winos   which are
proportional     to    the   electromagnetic  fine-structure  constant
$\alpha_{\rm   em}$~\cite{EWthreshold}.   Taking   CP violation   into
account, these additional contributions read
\begin{eqnarray}
(\Delta h_b/h_b)_{\rm em} &=& 
-\, \frac{\alpha_{\rm em}\mu^*M_2^*}{4\pi\,s_W^2}\,\Bigg[\,
|U^{\tilde{t}}_{L1}|^2\, I(m_{\tilde{t}_1}^2,|M_2|^2,|\mu|^2)\:
+\: |U^{\tilde{t}}_{L2}|^2\,I(m_{\tilde{t}_2}^2,|M_2|^2,|\mu|^2)
\nonumber \\
&&+\ \frac{1}{2}\, |U^{\tilde{b}}_{L1}|^2\, 
I(m_{\tilde{b}_1}^2,|M_2|^2,|\mu|^2)\: +\: 
\frac{1}{2}\, |U^{\tilde{b}}_{L2}|^2\, 
I(m_{\tilde{b}_2}^2,|M_2|^2,|\mu|^2)\, \Bigg]
\nonumber\\
&& -\ \frac{\alpha_{\rm em}\mu^*M_1^*}{12\pi c_W^2}\,\Bigg[\,
\frac{1}{3}\,I(m_{\tilde{b}_1}^2,m_{\tilde{b}_2}^2,|M_1|^2)
\: +\: \frac{1}{2}\, |U^{\tilde{b}}_{L1}|^2\,
I(m_{\tilde{b}_1}^2,|M_1|^2,|\mu|^2)\nonumber\\
&&+\, \frac{1}{2}\, |U^{\tilde{b}}_{L2}|^2\,
I(m_{\tilde{b}_2}^2,|M_1|^2,|\mu|^2)\: +\:
|U^{\tilde{b}}_{R1}|^2\, I(m_{\tilde{b}_1}^2,|M_1|^2,|\mu|^2)\nonumber\\
&&+\, |U^{\tilde{b}}_{R2}|^2\,I(m_{\tilde{b}_2}^2,|M_1|^2,|\mu|^2)\,\Bigg]\,, 
\end{eqnarray}
where 
\begin{equation}
  \label{Iabc}
I(a,b,c)\  =\ \frac{ab\, \ln  (a/b)\:  +\: bc\, \ln  (b/c)\:  + \:
ac\, \ln  (c/a)}{(a-b)\,(b-c)\,(a-c)}\, .
\end{equation}
We follow the  convention  of {\tt CPsuperH}~\cite{CPsuperH} for   the
mixing matrices of  the stops $U^{\tilde{t}}$, sbottoms $U^{\tilde{b}}$
and staus $U^{\tilde{\tau}}$.

There  are formulae analogous to  those above for the loop corrections
to the  $H_{1,2,3} t {\bar t}$ vertices,  which would  be relevant for
CP-violation measurements in $e^-  e^+   \to  \nu \bar{\nu}  t   {\bar
t}$~\cite{Marek}, for example.

Analogous   exchanges of  binos  and  winos  give  finite loop-induced
corrections  to  the $H_{1,2,3}\tau^+\tau^-$   coupling, which  have a
similar structure:
\begin{equation}
h_\tau\ =\ \frac{\sqrt{2}m_\tau}{v\cos\beta}\,
\frac{1}{1+(\Delta h_\tau/h_\tau)\tan\beta} \,,
\end{equation}
where
\begin{eqnarray}
(\Delta h_\tau/h_\tau) &=&
-\, \frac{\alpha_{\rm em}\,\mu^*M_2^*}{4\pi\,s^2_W}\,
\Bigg[\, I(m_{\tilde{\nu}_\tau}^2,|M_2|^2,|\mu|^2)\:
+\: \frac{1}{2}\, |U^{\tilde{\tau}}_{L1}|^2\,
I(m_{\tilde{\tau}_1}^2,|M_2|^2,|\mu|^2)\nonumber\\
&&+\, \frac{1}{2}\, |U^{\tilde{\tau}}_{L2}|^2\,
I(m_{\tilde{\tau}_2}^2,|M_2|^2,|\mu|^2)\,\Bigg]
\ +\ \frac{\alpha_{\rm em}\,\mu^*M_1^*}{4\pi\,c_W^2}\,
\Bigg[\, I(m_{\tilde{\tau}_1}^2,m_{\tilde{\tau}_2}^2,
|M_1|^2)\nonumber\\
&& +\: \frac{1}{2}\, |U^{\tilde{\tau}}_{L1}|^2\,
I(m_{\tilde{\tau}_1}^2,|M_1|^2,|\mu|^2)\: +\:
\frac{1}{2}\, 
|U^{\tilde{\tau}}_{L2}|^2\, I(m_{\tilde{\tau}_2}^2,|M_1|^2,|\mu|^2)
\nonumber \\
&&-\,|U^{\tilde{\tau}}_{R1}|^2\,I(m_{\tilde{\tau}_1}^2,|M_1|^2,|\mu|^2)
\: -\: |U^{\tilde{\tau}}_{R2}|^2\,
I(m_{\tilde{\tau}_2}^2,|M_1|^2,|\mu|^2)\Bigg]\,.
\end{eqnarray}
The threshold corrections modify the couplings of the neutral Higgs bosons to
the scalar and pseudoscalar fermion bilinears as follows~\cite{CEMPW}:
\begin{eqnarray}
\hspace{-1.0 cm}
g^S_{H_i\bar{f}f}&=&
\frac{O_{\phi_1i}}{\cos\beta}\ 
\real\left({\frac{1}{1+\kappa_f \tan\beta}}\right)\
+\ \frac{O_{\phi_2i}}{\cos\beta}\ \real\left({\frac{\kappa_f}{1+\kappa_f
\tan\beta}}\right)\nonumber\\ 
&& +\: O_{ai}\,
\imag\left({\frac{\kappa_f\,(\tan^2\beta+1)}{1+\kappa_f \tan\beta}}\right)\,,
\nonumber\\
g^P_{H_i\bar{f}f} &=&
\frac{O_{\phi_1i}}{\cos\beta}\
\imag\left({\frac{\kappa_f\,\tan\beta}{1+\kappa_f
\tan\beta}}\right)\ -\
\frac{O_{\phi_2i}}{\cos\beta}\ \imag\left({\frac{\kappa_f}{1+\kappa_f
\tan\beta}}\right)\nonumber\\ 
&&-\ O_{ai}\,
\real\left({\frac{\tan\beta-\kappa_f}{1+\kappa_f \tan\beta}}\right)\,,
\end{eqnarray}
where $f=b$ and $\tau^-$ and
\begin{equation}
\kappa_b\ =\ \frac{(\Delta h_b/h_b)}{1+(\delta h_b/h_b)}\,,\qquad
\kappa_\tau\ =\ (\Delta h_\tau/h_\tau)\,.
\end{equation}
There are similar formulae for the $H_{1,2,3}\mu^+\mu^-$ vertices that
would be  relevant    for $\mu^+  \mu^-$  colliders.   The   analogous
corrections to  the $H_{1,2,3}e^+e^-$ vertices   may be neglected.  

Additional  contributions  to  Higgs-boson  vertices  may  arise  from
absorptive  effects due  to the  opening of  third-generation sfermion
pair production channels. However, if the $H_{1,2,3}$-boson masses are
well below the kinematic  threshold of such production channels, these
absorptive effects are small and can be neglected.  Finally, we remind
the  reader  that  detailed  analytic expressions  for  the  effective
Higgs-boson couplings to the photon, the gluon, the $W^\pm$ bosons and
SUSY particles are given in~\cite{CPsuperH}.

\bigskip\bigskip

\setcounter{equation}{0}
\section{Tau Pair Production at the LHC}
\label{sec:production}

To       further     elucidate     the      formalism  presented    in
Section~\ref{sec:formalism},  we  now   discuss   in more detail   the
production, mixing  and decay of  Higgs bosons into  polarized $\tau^+
\tau^-$  pairs at the  LHC. We will  study individually the three most
significant production  channels for Higgs  bosons in the MSSM  at the
LHC:   (i) $b\bar{b}$ fusion,  (ii) $gg$ fusion and (iii)  $W^+W^-$
fusion.


\subsection{$b {\bar b}$ Fusion}

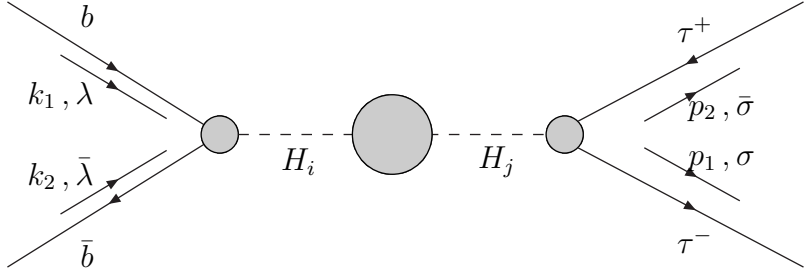
\begin{figure}[t]
\begin{center}
\begin{picture}(250,120)(0,20)

\ArrowLine(-30,100)(50,50)
\ArrowLine(50,50)(-30,0)
\DashCArc(50,50)(7,0,360){4}
\GOval(50,50)(7,7)(0){0.8}
\DashLine(57,50)(180,50){4}
\DashCArc(115,50)(15,0,360){4}
\GOval(115,50)(15,15)(0){0.8}
\DashCArc(180,50)(7,0,360){4}
\GOval(180,50)(7,7)(0){0.8}
\ArrowLine(270,100)(185,55)
\ArrowLine(185,45)(270,0)
\Text(0,95)[]{$b$}
\Text(0, 5)[]{$\bar{b}$}
\Text(80,40)[]{$H_i$}
\Text(155,40)[]{$H_j$}
\Text(230,90)[]{$\tau^+$}
\Text(230,10)[]{$\tau^-$}
\ArrowLine(210,55)(246,75)
\ArrowLine(210,45)(246,25)
\Text(240,60)[]{$p_2\,,\bar{\sigma}$}
\Text(240,40)[]{$p_1\,,\sigma$}
\ArrowLine(-10,80)(30,56)
\ArrowLine(-10,20)(30,44)
\Text(-10,35)[]{$k_2\,,\bar{\lambda}$}
\Text(-10,65)[]{$k_1\,,\lambda$}

\end{picture}\\
\end{center}
\smallskip
\noindent
\caption{\it Mechanisms contributing to the process $b\bar{b} \to
H\to \tau^+\tau^-$, including off-diagonal absorptive parts in the
Higgs-boson propagator matrix.}\label{f1}

\end{figure}

At large  $\tan\beta$, an  important  mechanism for producing  neutral
Higgs     bosons       at    the    LHC    is      $b     {\bar    b}$
fusion~\cite{BBH1,BBH2,GB2BH,BLS}.    Figure~\ref{f1} illustrates  how
the   matrix element ${\cal  M}^{b\bar{b}}$  for $b\bar{b} \rightarrow
\tau^+\tau^-$ receives contributions from the $s$-channel exchanges of
the neutral  Higgs  bosons.  The loop-corrected  propagator matrix and
vertices calculated  in the previous  section  are indicated by shaded
circles. The matrix element can be written as
\begin{eqnarray}
  \label{Mbb}
{\cal M}^{b\bar{b}}&=&-\frac{g^2 m_b m_\tau}
{4 M_W^2 \hat{s}}\sum_{i,j=1,2,3}
\ \sum_{\alpha ,\beta=\pm}
\Bigg\{(g^S_{H_i\bar{b}b}+i\alpha g^P_{H_i\bar{b}b})\,
\bar{v}(k_2,\bar{\lambda})P_\alpha u(k_1,\lambda)\: D_{ij}(\hat{s})\: 
\nonumber \\
&&\hspace{3.0 cm}
\times
(g^S_{H_j\tau^+\tau^-}+i \beta g^P_{H_j\tau^+\tau^-})\,
\bar{u}(p_1,\sigma)P_\beta v(p_2,\bar{\sigma})\Bigg\}\,,
\end{eqnarray}
where  $P_\alpha=(1+\alpha\gamma_5)/2$ and the running bottom-quark mass
at the scale of $\sqrt{\hat{s}}$ is used.
We  denote the helicities  of
$\tau^-$ and $\tau^+$ by $\sigma$ and $\bar{\sigma}$  and those of the
$b$ and $\bar b$ by $\lambda$  and $\bar{\lambda}$, respectively, with
$\sigma,\lambda=+$  and   $-$ standing   for  right- and   left-handed
particles. The four-momenta of the $\tau^-$ and $\tau^+$ are $p_1$ and
$p_2$, respectively, those of the $b$  and $\bar{b}$ are $k_1$ and
$k_2$,  respectively, and   $\hat{s}$  is  the  centre-of-mass  energy
squared of  the  ${\bar b} b$  pair  that  fuses into  a  Higgs boson:
$\hat{s}=(k_1+k_2)^2=(p_1+p_2)^2$. 

\vspace*{0.5cm}

An  important element  of our  formalism is  the consideration  of the
`full'  $3  \times   3$  Higgs-boson  propagator  matrix  $D(\hat{s})$
in~(\ref{Mbb})~\footnote{Strictly  speaking,  the complete  propagator
matrix $D(\hat{s})$ is a $4\times 4$-dimensional matrix spanned by the
basis~$(H_1,H_2,H_3,G^0)$~\cite{APNPB}.     However,    to   a    good
approximation,  we may neglect  the small  off-resonant self-energy
transitions of  the Higgs bosons  $H_{1,2,3}$ to the  neutral would-be
Goldstone boson $G^0$.}.  This is given by
\begin{equation}
  \label{eq:hprop}
D (\hat{s}) = \hat{s}\,
\left(\begin{array}{ccc}
       \hat{s}-M_{H_1}^2+i\imag\widehat{\Pi}_{11}(\hat{s}) &
i\imag\widehat{\Pi}_{12}(\hat{s})&
       i\imag\widehat{\Pi}_{13}(\hat{s}) \\
       i\imag\widehat{\Pi}_{21}(\hat{s}) &
\hat{s}-M_{H_2}^2+i\imag\widehat{\Pi}_{22}(\hat{s})&
       i\imag\widehat{\Pi}_{23}(\hat{s}) \\
       i\imag\widehat{\Pi}_{31}(\hat{s}) & i\imag\widehat{\Pi}_{32}(\hat{s}) &
       \hat{s}-M_{H_3}^2+ 
       i\imag\widehat{\Pi}_{33}(\hat{s})
      \end{array}\right)^{-1} \,,
\end{equation}
where the inversion of the 3-by-3 matrix  is carried out analytically.
In (\ref{eq:hprop}), the absorptive  parts of the Higgs  self-energies
$\imag      \widehat{\Pi}_{ij}(\hat{s})$          are    given      in
Section~\ref{sec:formalism}    and  $M_{H_{1,2,3}}$ are  the  one-loop
Higgs-boson pole masses,    where higher-order absorptive  effects  on
$M_{H_{1,2,3}}$ have been  ignored~\cite{CEPW2}.  In the same context,
the off-shell dispersive parts of the Higgs-boson self-energies in the
Higgs-boson propagator matrix $D  (\hat{s})$ have also been neglected,
since  these are formally higher-order  effects and  very small in the
relevant Higgs-boson     resonant   region.   Finally,    we   include
in~(\ref{Mbb}) the finite loop-induced corrections to the couplings of
Higgs  bosons to $b$     quarks, $g^{S,P}_{H_i\bar{b}b}$, and   $\tau$
leptons, $g^{S,P}_{H_j\tau^+\tau^-}$, due to the exchanges of gauginos
and Higgsinos, as has been discussed in Section~\ref{sec:formalism}.

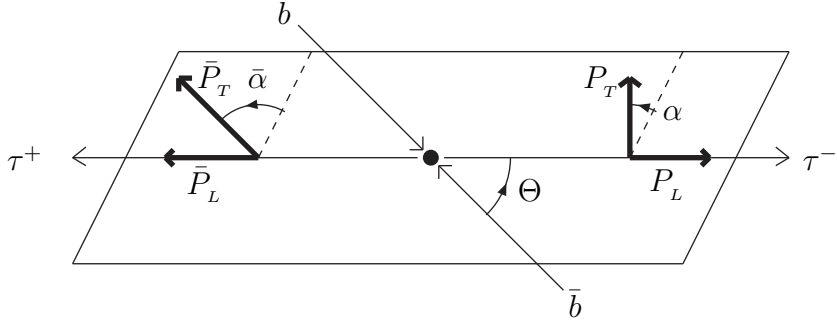
\begin{figure}[t]
\bigskip
\begin{center}
\begin{picture}(300,100)(0,0)

\Line(15,10)(55,90)
\Line(55,90)(285,90)
\Line(285,90)(245,10)
\Line(245,10)(15,10)

\Text(-3,50)[]{$\tau^+$}
\Line(15,50)(145,50)
\Line(15,50)(20,54)
\Line(15,50)(20,47)

\Text(298,50)[]{$\tau^-$}
\Line(155,50)(285,50)
\Line(285,50)(280,54)
\Line(285,50)(280,47)

\Vertex(150,50){3}

\Text(85,80)[]{$\bar{\alpha}$}
\ArrowArc(85,50)(20,60,135)
\Text(242,67)[]{$\alpha$}
\ArrowArc(225,50)(20,60,90)

\Line(100,100)(147,53)
\Line(147,53)(147,58)
\Line(147,53)(142,53)
\Text(95,105)[]{$b$}

\Line(200,0)(153,47)
\Line(153,47)(153,42)
\Line(153,47)(158,47)
\Text(205,-5)[]{$\bar{b}$}

\ArrowArc(150,50)(30,314,360)
\Text(188,38)[]{$\Theta$}

\DashLine(85,50)(105,90){3}
\DashLine(225,50)(245,90){3}

\SetWidth{2}
\Line(225,50)(225,80)
\Line(225,80)(221,77)
\Line(225,80)(229,77)
\Text(69,80)[]{$\bar{P}_{_T}$}

\Line(225,50)(255,50)
\Line(255,50)(252,53)
\Line(255,50)(252,47)
\Text(240,40)[]{${P}_{_L}$}

\Line(85,50)(55,80)
\Line(55,80)(60,80)
\Line(55,80)(55,75)
\Text(215,79)[]{$P_{_T}$}

\Line(85,50)(50,50)
\Line(50,50)(53,53)
\Line(50,50)(53,47)
\Text(65,40)[]{$\bar{P}_{_L}$}

\end{picture}\\
\end{center}
\bigskip
\noindent
\caption{\it The $\tau^+ \tau^-$ production plane with definitions 
            of the scattering angle $\Theta$. The
            transverse polarization vectors $P_T$ and $\bar{P}_T$
            have azimuthal angles $\alpha$ and $\bar{\alpha}$, 
respectively, with respect to the event plane.}\label{f2}
\end{figure}

In the centre-of-mass coordinate system for the $b {\bar b}$ pair, the 
helicity amplitudes are given by
\begin{equation}
{\cal M}^{b\bar{b}}(\sigma\bar{\sigma};\lambda\bar{\lambda})\ =\
-\frac{g^2 m_b m_\tau}{4 M_W^2}\,
\langle \sigma ; \lambda \rangle_{b}
\delta_{\sigma\bar{\sigma}}\delta_{\lambda\bar{\lambda}}\,,
\label{eq:bbamp}
\end{equation}
where
\begin{equation}
\langle \sigma ; \lambda \rangle_{b}\  \equiv\ \sum_{i,j=1,2,3}
(\lambda\beta_b\,g^S_{H_i\bar{b}b}+i g^P_{H_i\bar{b}b})\: D_{ij}(\hat{s})\:
(\sigma\beta_\tau\,g^S_{H_j\tau^+\tau^-}-i g^P_{H_j\tau^+\tau^-})\,,
\end{equation}
with  $\beta_f=\sqrt{1-4m_f^2/\hat{s}}$.  Note that the cross sections
for general (longitudinal or  transverse) $\tau^\pm$ polarizations can
be   computed from     the    helicity  amplitudes   by   a   suitable
rotation~\cite{HaZe} from the helicity basis to a general spin basis.

The $\tau$-polarization weighted squared matrix elements are given by
\begin{equation}
\overline{\left|{\cal M}^{b\bar{b}}\right|^2}\ 
=\ \frac{1}{12}\, \sum_{\lambda=\pm} \left(
\sum_{\sigma\sigma^\prime\bar{\sigma}\bar{\sigma}^\prime} 
{\cal M}^{b\bar{b}}_{\sigma\bar{\sigma}}
{\cal M}^{b\bar{b}*}_{\sigma^\prime\bar{\sigma}^\prime}
\bar{\rho}_{\bar{\sigma}^\prime\bar{\sigma}}
\rho_{\sigma^\prime\sigma}
\right)\ =\ \frac{1}{12}\, \sum_{\lambda=\pm} 
{\rm Tr}\left[{\cal M}^{b\bar{b}}\bar{\rho}^T
{\cal M}^{b\bar{b}\dagger}\rho\right]\,,
\label{eq:bsq}
\end{equation}
where $\rho$ and $\bar{\rho}$ are $2\times 2$ polarization density
matrices for the $\tau^-$ and $\tau^+$, respectively:
\begin{equation}
\rho\ =\ \frac{1}{2}\left(\begin{array}{cc}
     1+P_L                  & P_T\,{\rm e}^{-i\alpha}  \\[1mm]
    P_T\,{\rm e}^{i\alpha} & 1-P_L
                   \end{array}\right)\,,\qquad
\bar{\rho}\ =\ \frac{1}{2}\left(\begin{array}{cc}
     1+\bar{P}_L       & -\bar{P}_T\,{\rm e}^{i\bar{\alpha}}  \\[1mm]
    -\bar{P}_T\,{\rm e}^{-i\bar{\alpha}} & 1-\bar{P}_L
                   \end{array}\right)\,.
\end{equation}
Evaluating the trace in (\ref{eq:bsq}) yields
\begin{eqnarray}
\overline{\left|{\cal M}^{b\bar{b}}\right|^2} 
&=&\frac{1}{12}
\left( \frac{g^2m_bm_\tau}{4 M_W^2} \right)^2
\Bigg\{ C^b_1 (1+P_L\bar{P}_L)
+C^b_2 (P_L+\bar{P}_L) \nonumber \\
&& \hspace{3.3 cm}
+P_T\bar{P}_T\left[C^b_3\cos(\alpha-\bar{\alpha})
+C^b_4\sin(\alpha-\bar{\alpha})\right]
\Bigg\}\,.
\label{Msquared}
\end{eqnarray}
The $\tau^+ \tau^-$  production   plane is depicted   schematically in
Fig.~\ref{f2}, where  the transverse  polarization angles $\alpha$ and
$\bar{\alpha}$ are also defined.

The coefficients $C^b_n(n=1-4)$  in  (\ref{Msquared}) are  defined  in
terms of the helicity amplitudes by
\begin{eqnarray}
  \label{Cbn}
C^b_1 \!\!&\equiv &\!\! \frac{1}{4}\sum_{\lambda=\pm}\left(
|\langle +;\lambda\rangle_{b}|^2 +|\langle -;\lambda\rangle_{b}|^2
\right) \,, \qquad
C^b_2\ \equiv\ \frac{1}{4}\sum_{\lambda=\pm}\left(
|\langle +;\lambda\rangle_{b}|^2 -|\langle -;\lambda\rangle_{b}|^2
\right) \,,  \nonumber \\
C^b_3 \!\!&\equiv &\!\! -\frac{1}{2}\sum_{\lambda=\pm}\real\left(
\langle +;\lambda\rangle_{b} \langle -;\lambda\rangle_{b}^*
\right) \,,  \qquad\quad\,
C^b_4\ \equiv\ \frac{1}{2}\sum_{\lambda=\pm}\imag\left(
\langle +;\lambda\rangle_{b} \langle -;\lambda\rangle_{b}^*
\right) \,.
\label{eq:c}
\end{eqnarray}
Under CP   and CP$\widetilde{\rm T}$~\footnote{We define $\widetilde{\rm
T}$ as the  naive  ${\rm T}$-reversal transformation, under  which the
spins and  3-momenta of  the  asymptotic states reverse sign,  without
interchanging initial  to  final   states.  In  addition,    under the
operation    of $\widetilde{\rm T}$,     the  matrix element gets  complex
conjugated.}   transformations, the helicity  amplitudes  transform as
follows:
\begin{equation}
\langle \sigma;\lambda\rangle_{b}
\, \stackrel{\rm CP}{\leftrightarrow}  \,
+\langle -\sigma;-\lambda\rangle_{b}\,, \qquad
\langle \sigma;\lambda\rangle_{b}
\, \stackrel{\rm CP\widetilde{\rm T}}{\leftrightarrow} \,
+\langle -\sigma;-\lambda\rangle_{b}^*\,.
\end{equation}
Hence, the  CP  and CP$\widetilde{\rm  T}$  parities of   the coefficients
$C^b_n$ defined in (\ref{Cbn}) are given by
\begin{equation}
  \label{CPparities}
C^b_1[++]\,, \qquad C^b_2[--]\,, \qquad
C^b_3[++]\,, \qquad C^b_4[-+]\,,
\end{equation}
where the first  and second symbols in  the square brackets are the CP
and  CP$\widetilde{\rm  T}$ parities, respectively.  Consequently, the
coefficients  $C^b_2$  and  $C^b_4$ signify genuine   phenomena  of CP
violation,  whereas $C^b_1$ and  $C^b_3$  are CP-conserving.  Here, we
should remark that a non-zero  value for the CP$\widetilde{\rm T}$-odd
coefficient  $C^b_2$ can only  be  induced by non-vanishing absorptive
effects.    In our case,  such   effects  mainly  originate from   the
absorptive parts of the Higgs-boson self-energies.

Finally,  for          our    phenomenological         discussion   in
Section~\ref{sec:numerical}, we define the parton-level cross sections
\begin{equation}
  \label{eq:bparton}
\hat{\sigma}_i(b\bar{b}\to H \to \tau^+\tau^-)\ \equiv\
\frac{\beta_\tau}{192\pi \hat{s}}\
\left(\frac{g^2m_bm_\tau}{4M_W^2}\right)^2\,C^b_i\,,
\end{equation}
where the intermediate state  $H$ collectively denotes all the $H_i\to
H_j$ resonant transitions with $i,j = 1,2,3$.

\subsection{$gg$ Fusion}
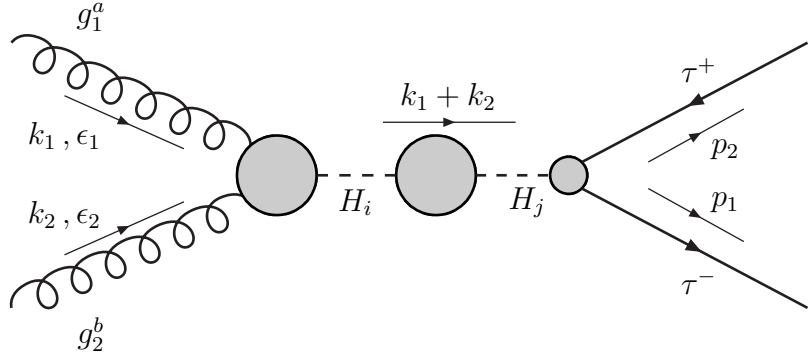
\begin{figure}[t]
\vspace*{1cm}
\begin{center}

\SetWidth{1.0}

\begin{picture}(250,100)(0,0)
\Gluon(-30,100)(60,60){6}{6}
\Gluon(-30,0)(60,40){6}{6}
\DashCArc(70,50)(15,0,360){4}
\GOval(70,50)(15,15)(0){0.8}
\DashCArc(130,50)(15,0,360){4}
\GOval(130,50)(15,15)(0){0.8}
\DashLine(85,50)(115,50){4}
\DashLine(145,50)(180,50){4}
\DashCArc(180,50)(7,0,360){4}
\GOval(180,50)(7,7)(0){0.8}
\ArrowLine(270,100)(185,55)
\ArrowLine(185,45)(270,0)
\Text(0,110)[]{$g_1^a$}
\Text(0,-10)[]{$g_2^b$}
\Text(100,40)[]{$H_i$}
\Text(165,40)[]{$H_j$}
\Text(230,90)[]{$\tau^+$}
\Text(230,10)[]{$\tau^-$}

\SetWidth{0.5}

\ArrowLine(210,55)(246,75)
\ArrowLine(210,45)(246,25)
\Text(240,60)[]{$p_2$}
\Text(240,40)[]{$p_1$}
\ArrowLine(-10,20)(35,40)
\ArrowLine(-10,80)(35,60)
\Text(-10,35)[]{$k_2\,,\epsilon_2$}
\Text(-10,65)[]{$k_1\,,\epsilon_1$}
\ArrowLine(110,70)(160,70)
\Text(135,80)[]{$k_1+k_2$}
\end{picture}
\end{center}
\smallskip
\noindent
\caption{\it Mechanisms contributing to the process
            $gg \to H \to \tau^+\tau^-$ via the three 
            neutral Higgs bosons $H_{1,2,3}$.}\label{f3}
\end{figure}

The matrix element  ${\cal M}^{gg}$ for the process  $gg \rightarrow H
\to \tau^+\tau^-$, depicted in Fig.~\ref{f3}, can be written as
\begin{eqnarray}
{\cal M}^{gg} &=& \frac{g\alpha_s m_\tau\delta^{ab}}{8\pi v
M_W}\nonumber\\
&&\hspace{-1cm}\times\, \sum_{i,j=1,2,3}^{3}\ \sum_{\alpha=\pm}
G_{H_i}(k_1,\epsilon_1;k_2,\epsilon_2)\: D_{ij}(\hat{s})\:
(g_{H_j\tau^+\tau^-}^S+i \alpha
g_{H_j\tau^+\tau^-}^P)\: \bar{u}(p_1,\sigma) P_\alpha
v(p_2,\bar{\sigma})\,.\qquad\quad
\end{eqnarray}
In the above, $a$ and $b$ are indices
of the  SU(3) generators in the  adjoint representation  and $k_{1,2}$
and $\epsilon_{1,2}$ are the four-momenta and  wave vectors of the two
gluons, respectively.  Again, we denote the helicities of $\tau^-$ and
$\tau^+$  by $\sigma$  and   $\bar{\sigma}$  with $\sigma=+$ and   $-$
standing for  right-  and left-handed particles.   The four-momenta of
$\tau^-$   and $\tau^+$  are    $p_1$  and $p_2$,  respectively,   and
$\hat{s}=(k_1+k_2)^2=(p_1+p_2)^2$.   The Higgs-boson propagator matrix
$D(\hat{s})$   was  given  in (\ref{eq:hprop})  and   the loop-induced
couplings of the Higgs bosons $H_i$ to two gluons are given by
\begin{eqnarray}
G_{H_i}(k_1,\epsilon_1;k_2,\epsilon_2) &=& i\,S^g_i(\sqrt{\hat{s}})
\left(\epsilon_1\cdot\epsilon_2\: -\: \frac{2}{\hat{s}}\,
k_1\cdot\epsilon_2 k_2\cdot\epsilon_1\right)\ -\ i\,
P^g_i(\sqrt{\hat{s}})\ \frac{2}{\hat{s}}\,
\varepsilon_{\mu\nu\rho\sigma}\epsilon_1^\mu\epsilon_2^\nu    k_1^\rho
k_2^\sigma\,, \quad\nonumber\\
\end{eqnarray}
with  $\varepsilon_{0123}  = 1$. For the    loop functions $S^g_i$ and
$P^g_i$, we follow the definitions of~\cite{CPsuperH}.

In  the two-gluon  centre-of-mass  coordinate system with  ${\bf k}_1$
along the  positive $z$ direction and  ${\bf k}_2$ along  the negative
$z$ direction, the wave vectors of two photons are given by
\begin{equation}
\epsilon^\mu_1(\lambda_1)\ =\ \frac{1}{\sqrt{2}}\, \Big(0,-\lambda_1,
-i,0\Big)\,,\qquad
\epsilon^\mu_2(\lambda_2)\ =\ \frac{1}{\sqrt{2}}\, 
\Big(0,-\lambda_2, i,0\Big)\,,
\end{equation}
where $\lambda=+1$ and $-1$
denote the right and left gluon helicities, respectively.
The helicity amplitudes are given by
\begin{equation}
{\cal M}^{gg}(\sigma\bar{\sigma};\lambda_1\lambda_2)\ =\ 
\frac{g\alpha_s m_\tau\sqrt{\hat{s}}\delta^{ab}}{8\pi v M_W}
\langle\sigma;\lambda_1\rangle_g
\delta_{\sigma\bar{\sigma}}\delta_{\lambda_1\lambda_2}\,, 
\label{eq:ggamp}
\end{equation}
where the amplitude $\langle\sigma;\lambda\rangle_g$ is defined as
\begin{equation}
\langle\sigma;\lambda\rangle_g \equiv \sum_{i,j=1,2,3}
[S_i^g(\sqrt{\hat{s}})+i\lambda P_i^g(\sqrt{\hat{s}})]\:
D_{ij}(\hat{s})\: (\sigma\beta_\tau g_{H_j\tau^+\tau^-}^S -i
g_{H_j\tau^+\tau^-}^P)\,.
\end{equation}
We note that the amplitude (\ref{eq:ggamp}) has the same structure
as the amplitude (\ref{eq:bbamp}) for $b\bar{b}\rightarrow \tau^+\tau^-$,
except for the overall constant. We obtain from the helicity 
amplitudes the polarization-weighted squared matrix elements given by
\begin{eqnarray}
\overline{\left|{\cal M}^{gg}\right|^2} 
&=&\frac{1}{32}
\left( \frac{g\alpha_s m_\tau\sqrt{\hat{s}}}{8\pi v M_W} \right)^2
\Bigg\{ C^g_1 (1+P_L\bar{P}_L)
+C^g_2 (P_L+\bar{P}_L) \nonumber \\
&& \hspace{3.3 cm}
+P_T\bar{P}_T\left[C^g_3\cos(\alpha-\bar{\alpha})
+C^g_4\sin(\alpha-\bar{\alpha})\right]
\Bigg\}\,,
\end{eqnarray}
where the coefficients $C_n^g$ are obtained by replacing 
$\langle\sigma;\lambda\rangle_b \to \langle\sigma;\lambda\rangle_g$
and interpreting $\lambda$ as the gluon helicity in (\ref{eq:c}).

Under the CP and CP$\widetilde{\rm T}$ transformations, the helicity 
amplitudes transform as follows:
\begin{equation}
\langle \sigma;\lambda\rangle_{g}
\, \stackrel{\rm CP}{\leftrightarrow}  \,
-\langle -\sigma;-\lambda\rangle_{g}\,, \qquad
\langle \sigma;\lambda\rangle_{g}
\, \stackrel{\rm CP\widetilde{\rm T}}{\leftrightarrow} \,
-\langle -\sigma;-\lambda\rangle_{g}^*\,,
\end{equation}
where the CP and CP$\widetilde{\rm T}$ 
parities of the coefficients $C^g_i$ are the same as those of the $C^b_i$.
Finally, we define the parton-level cross sections as:
\begin{equation}
\hat{\sigma}_i(gg\to H\to \tau^+\tau^-)\ \equiv\ 
\frac{\beta_\tau}{512\pi \hat{s}}
\left(\frac{g\alpha_s m_\tau\sqrt{\hat{s}}}{8\pi v M_W}\right)^2\,C^g_i\,.
\label{eq:gparton}
\end{equation}
Note  that  the  CP-   and  CP$\widetilde{\rm  T}$-odd  cross  section
$\hat{\sigma}_2$ receives  contributions from the  absorptive parts of
the $H_{1,2,3}gg$ vertices and Higgs-boson self-energies as well.

\subsection{$W^+ W^-$ Fusion}

The last  important mechanism for  the production of the  MSSM neutral
Higgs    bosons     at    the     LHC    is    $W^+     W^-$    fusion
\cite{WWF0,WWF1,WWF2,WWTAUTAU}.   The matrix  element  ${\cal M}^{WW}$
for this process, $W^-(k_1)  W^+(k_2) \rightarrow H \rightarrow \tau^-
(p_1)\tau^+ (p_2)$ with $\hat{s}=(k_1+k_2)^2$, is given by
\begin{equation}
{\cal M}^{WW}\ =\ \frac{g^2 m_\tau}{2\hat{s}}
\sum_{i,j=1}^3\ \sum_{\alpha =\pm}
g_{H_iVV} \,\epsilon_1 \cdot \epsilon_2\:
D_{ij}(\hat{s})\:
(g_{H_j\tau^+\tau^-}^S+i \alpha g_{H_j\tau^+\tau^-}^P)
\, \bar{u}(p_1,\sigma) P_\alpha v(p_2,\bar{\sigma})\,,
\end{equation}
where  $\epsilon_1$ and $\epsilon_2$ are   the polarization vectors of
two vector bosons and  $g_{H_iVV}$ denotes the  coupling of the  Higgs
boson  $H_i$ with a pair   of  gauge bosons,  as  defined through  the
interaction Lagrangian
\begin{equation}
{\cal L}_{HVV}\ =\ g M_W \left(W^+_\mu W^{-\mu}+\frac{1}{2 c_W^2} Z_\mu
Z^\mu\right)\: \sum_{i=1}^3\, g_{H_iVV} H_i\,.
\end{equation}
In the  $W^+  W^-$ centre-of-mass  coordinate system  with ${\bf k}_1$
along the  positive $z$ direction  and ${\bf k}_2$ along  the negative
$z$ direction, the polarization vectors of two vector bosons are given
by
\begin{eqnarray}
\epsilon_1^\mu(\lambda_1=\pm 1)\!\!&=&\!\! 
\frac{1}{\sqrt{2}}\,\Big(0,\mp 1,-i,0\Big)\,, \qquad\ 
\epsilon_1^\mu(\lambda_1=0)\ =\ \frac{1}{\sqrt{k_1^2}}\,
\Big(|{\bf k}_1|,0,0,k^0_1\Big)\,, \nonumber \\
\epsilon_2^\mu(\lambda_2=\pm 1)
\!\!&=&\!\!\frac{1}{\sqrt{2}}\,\Big(0,\mp 1,i,0\Big)\,,\qquad\quad
\epsilon_2^\mu(\lambda_2=0)\ =\ \frac{1}{\sqrt{k_2^2}}\,
\Big(|{\bf k}_2|,0,0,-k^0_2\Big)\,, 
\end{eqnarray}
where     the    polarization    vectors      are  normalized       by
$\epsilon_i(\lambda)\cdot\epsilon_i^*(\lambda^\prime)=
-\delta_{\lambda\lambda^\prime}$, and  $\lambda=\pm 1$ and $\lambda=0$
denote the transverse (right  and  left helicities) and   longitudinal
polarizations, respectively. In this  frame, the helicity amplitude is
given by
\begin{equation}
{\cal M}^{WW}(\sigma\bar{\sigma};\lambda_1\lambda_2)\ =\
\frac{g^2m_\tau}{2\sqrt{\hat{s}}} \langle \sigma;\lambda_1\rangle_W
\delta_{\sigma\bar{\sigma}}\delta_{\lambda_1\lambda_2}\,,
\end{equation}
where the amplitude $\langle \sigma;\lambda\rangle_W$ is defined by
\begin{equation}
\langle \sigma;\lambda\rangle_W\ \equiv\ 
\sum_{i,j=1,2,3} \omega(\lambda)\,g_{H_iVV}\: D_{ij}(\hat{s})\:
(\sigma\beta_\tau\,g^S_{H_j\tau^+\tau^-}-i g^P_{H_j\tau^+\tau^-})\,,
\end{equation}
with 
\begin{equation}
\omega(\pm)=1\quad \mbox{and} \quad 
\omega(0)=-k_1\cdot k_2/\sqrt{k_1^2k_2^2} \,.
\end{equation}
The factor $\omega(0)$  becomes $1-\hat{s}/2M_W^2$ for on-shell vector
bosons and dominates the amplitude for $\hat{s}\gg M_W^2$.

\medskip

One can then obtain the following averaged amplitude squared:
\begin{eqnarray}
\overline{\left|{\cal M}^{WW}\right|^2} 
&=&\frac{1}{9}
\left( \frac{g^2m_\tau}{2\sqrt{\hat{s}}} \right)^2
\Bigg\{ C^W_1 (1+P_L\bar{P}_L)
+C^W_2 (P_L+\bar{P}_L) \nonumber \\
&& \hspace{2.3 cm}
+P_T\bar{P}_T\left[C^W_3\cos(\alpha-\bar{\alpha})
+C^W_4\sin(\alpha-\bar{\alpha})\right]
\Bigg\}\,,
\end{eqnarray}
where    the   coefficients $C_n^W$ can   be    obtained  by replacing
$\langle\sigma;\lambda\rangle_b  \to   \langle\sigma;\lambda\rangle_W$
and   summing  over   $\lambda=\pm,0$  in (\ref{eq:c}).    The CP  and
CP$\widetilde{\rm T}$   parities of the  coefficients  $C^W_n$ are the
same   as  those of $C^b_n$   or  $C^g_n$, and  the parton-level cross
sections are defined similarly as
\begin{equation}
  \label{eq:Wparton}
\hat{\sigma}_i(W^+_{T,L}W^-_{T,L}\to H \to \tau^+\tau^-)\
\equiv\ \frac{\beta_\tau}{144\, \pi\hat{s}}
\left( \frac{g^2m_\tau}{2\sqrt{\hat{s}}} \right)^2 C^W_i\,.
\end{equation}
In   kinematic   situations   where  the   longitudinal   $W^+_LW^-_L$
contributions can  be neglected, the average factor  $1/144$ should be
replaced by  $1/64$.  Finally, we  note that it is  straightforward to
calculate $ZZ$-fusion  processes in a similar  fashion, although their
cross  sections  are smaller  approximately  by  a  factor of  4  than
$W^+W^-$ collisions at the hadron level.

\setcounter{equation}{0}
\section{Numerical Examples}
\label{sec:numerical}

We now present some numerical examples of CP-violating Higgs signatures in
$\tau^+ \tau^-$ production at the LHC. As already mentioned, these
signatures may be enhanced at large $\tan \beta$, and three-way mixing is
potentially important for small charged Higgs-boson masses. Since the 
prospects for observing $H_{1,2,3}
\to \tau^+ \tau^-$ at the LHC are best for light Higgs bosons, we present
in this section some numerical analyses in a specific scenario in which
all the three Higgs states mix significantly. 

Explicitly, we take the following parameter set:
\begin{eqnarray}
  \label{MSSM1}
&&\tan\beta=50, \ \ M_{H^\pm}^{\rm pole}=155~~{\rm GeV}, 
\nonumber \\
&&M_{\tilde{Q}_3} = M_{\tilde{U}_3} = M_{\tilde{D}_3} =
M_{\tilde{L}_3} = M_{\tilde{E}_3} = M_{\rm SUSY} = 0.5 ~~{\rm TeV},
\nonumber \\
&& |\mu|=0.5 ~~{\rm TeV}, \ \
|A_{t,b,\tau}|=1 ~~{\rm TeV},   \ \
|M_2|=|M_1|=0.3~~{\rm TeV}, \ \ |M_3|=1 ~~{\rm TeV},
\nonumber \\
&&
\Phi_\mu = 0^\circ, \ \
\Phi_A=\Phi_{A_t} = \Phi_{A_b} = \Phi_{A_\tau} = 90^\circ, \ \
\Phi_1 = \Phi_2 = 0^\circ,
\end{eqnarray}
and we consider two values for the phase of the gluino mass parameter 
$M_3$: $\Phi_3 = -90^\circ\,, -10^\circ$. 
For $\Phi_3 = -10^\circ$, {\tt CPsuperH} yields for
the masses and widths of the neutral Higgs bosons: \\
\begin{eqnarray}
&&
M_{H_1}=120.2~~{\rm GeV}, \ \
M_{H_2}=121.4~~{\rm GeV}, \ \
M_{H_3}=124.5~~{\rm GeV}, \ \
\nonumber \\  &&
\Gamma_{H_1}=1.19~~{\rm GeV}, \ \ \ \
\Gamma_{H_2}=3.42~~{\rm GeV}, \ \ \ \ \ \
\Gamma_{H_3}=3.20~~{\rm GeV},
\end{eqnarray}
and for $\Phi_3 = -90^\circ$:
\begin{eqnarray}
&&
M_{H_1}=118.4~~{\rm GeV}, \ \
M_{H_2}=119.0~~{\rm GeV}, \ \
M_{H_3}=122.5~~{\rm GeV}, \ \
\nonumber \\  &&
\Gamma_{H_1}=3.91~~{\rm GeV}, \ \ \ \
\Gamma_{H_2}=6.02~~{\rm GeV}, \ \ \ \ \ \
\Gamma_{H_3}=6.34~~{\rm GeV},
\end{eqnarray}
respectively.

\begin{figure}[p]
\epsfig{figure=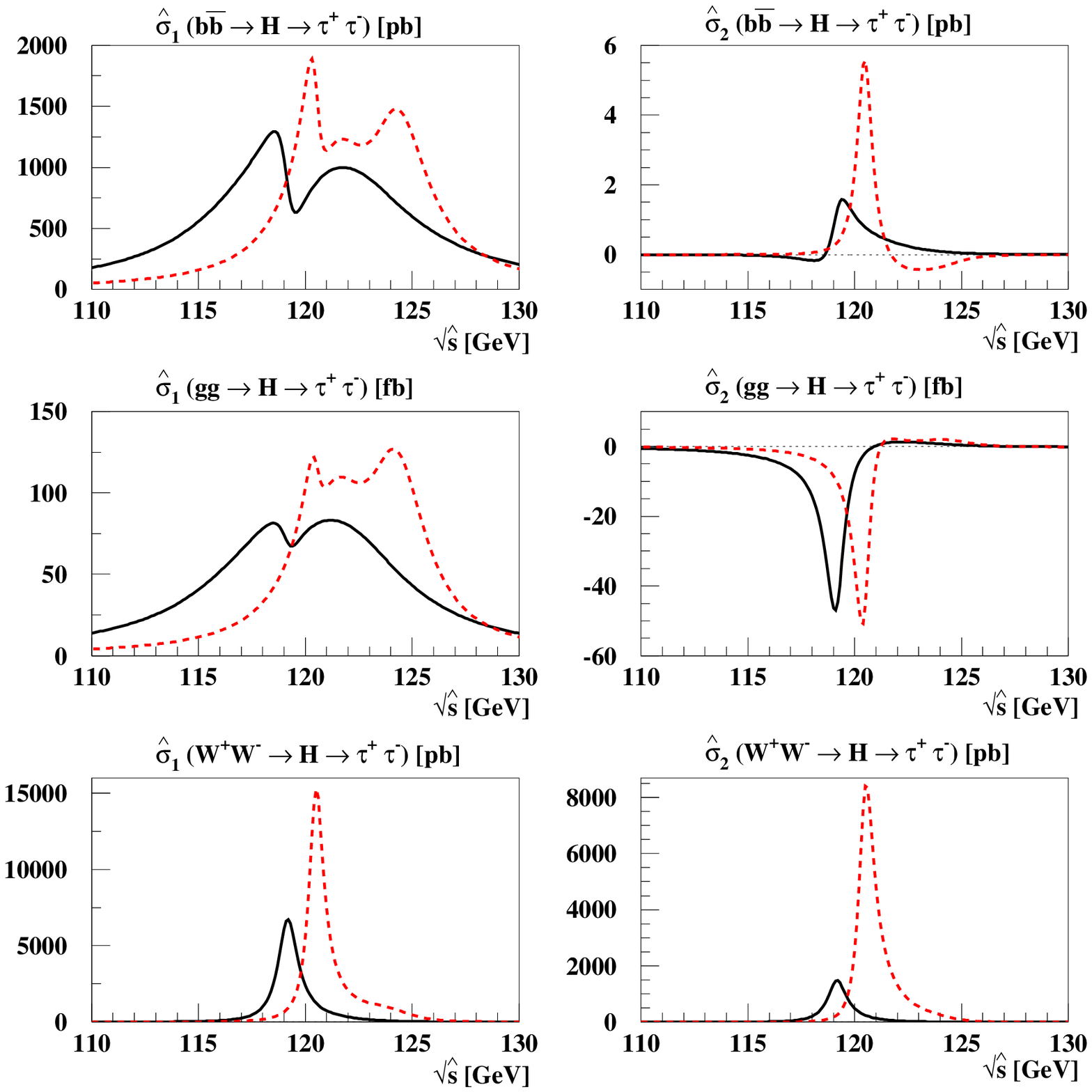,height=18cm,width=18cm}
\caption{\it The parton-level cross sections
$\hat{\sigma}_{1,2}(b\bar{b}\rightarrow H \rightarrow \tau^+\tau^-)$
in pb, $\hat{\sigma}_{1,2}(gg\rightarrow H \rightarrow \tau^+\tau^-)$
in fb, and $\hat{\sigma}_{1,2}(W^+W^-\rightarrow H \rightarrow
\tau^+\tau^-)$ in pb as functions of $\sqrt{{\hat s}}$.  The solid
lines are for the three-Higgs mixing scenario with $\Phi_3=-90^\circ$
and the dashed ones with $\Phi_3=-10^\circ$.}\label{fig:3mix}
\end{figure}

\begin{figure}[p]
\epsfig{figure=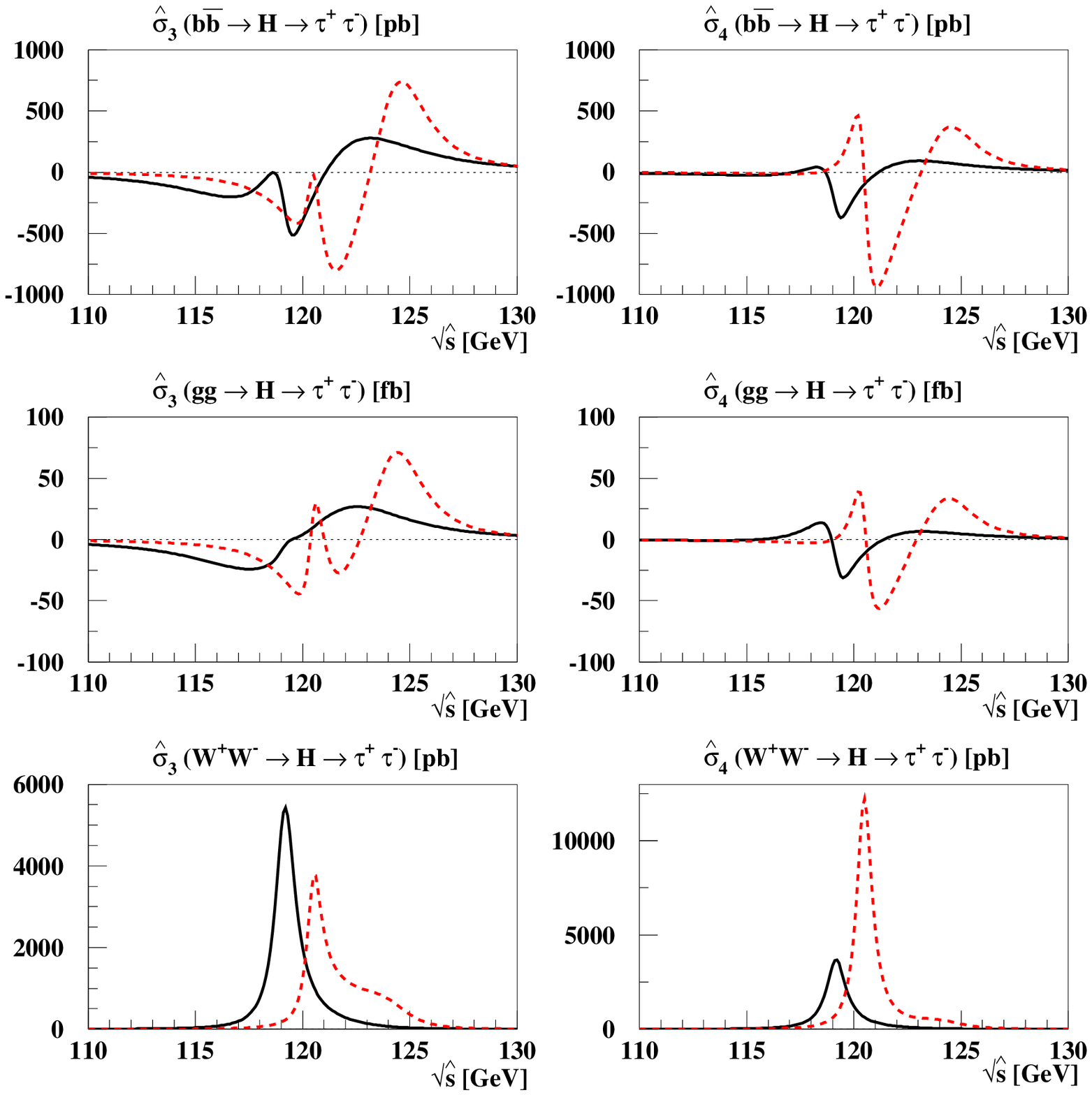,height=18cm,width=18cm}
\caption{\it The parton-level cross sections
$\hat{\sigma}_{3,4}(b\bar{b}\rightarrow H \rightarrow \tau^+\tau^-)$ in pb,
$\hat{\sigma}_{3,4}(gg\rightarrow H \rightarrow \tau^+\tau^-)$ in fb, and
$\hat{\sigma}_{3,4}(W^+W^-\rightarrow H \rightarrow \tau^+\tau^-)$ in pb
as functions of $\sqrt{{\hat s}}$.
The solid lines are for the three-Higgs mixing scenario
with $\Phi_3=-90^\circ$ and
the dashed ones with $\Phi_3=-10^\circ$.}
\label{fig:3mix34}
\end{figure}

In Figs.~\ref{fig:3mix} and~\ref{fig:3mix34}, we show the parton-level
cross sections $\hat{\sigma}_{i}(b\bar{b}\rightarrow  H \rightarrow  
\tau^+\tau^-)$,
$\hat{\sigma}_{i}(gg\rightarrow H \rightarrow \tau^+\tau^-)$    and
$\hat{\sigma}_{i}(WW\rightarrow  H \rightarrow \tau^+\tau^-)$ defined in
(\ref{eq:bparton}),   (\ref{eq:gparton})   and     (\ref{eq:Wparton}),
respectively, as functions   of  the $\tau^+ \tau^-$   invariant  mass
$\sqrt{{\hat s}}$.  The solid lines are for $\Phi_3=-90^\circ$ and the
dashed (red) ones for  $\Phi_3=-10^\circ$.  We recall that 
non-vanishing
of $\hat{\sigma}_2$ and $\hat{\sigma}_{4}$ are  direct signals of  CP
violation   in longitudinally  and   transversally   polarized
$\tau^+\tau^-$ pairs, respectively.

The   parton-level   cross  sections   $\hat{\sigma}_{i}(WW\rightarrow
\tau^+\tau^-)$ have  been computed  by neglecting the  contribution of
the longitudinally-polarized  $W^\pm$, i.e., setting $\omega(0)=0$.  For 
the
MSSM   scenario  defined   in~(\ref{MSSM1}),  this   is   a  plausible
approximation  for  Higgs-boson   masses  below  the  $WW$  threshold.
Possible  uncertainties that  such a  treatment may  introduce 
largely  cancel when  we consider  ratios  of the
cross sections $\hat{\sigma}_{i}(WW\rightarrow \tau^+\tau^-)$, such as
the CP asymmetries to be defined later in this section.

In   Fig.~\ref{fig:3mix},   we   observe   that  the   cross   section
$\hat{\sigma}_2$, which  quantifies CP violation in  the production of
longitudinally polarized  $\tau$-lepton pairs at the  parton level, is
comparable to the spin-averaged cross section $\hat{\sigma}_1$ in $WW$
and $gg$ collisions. This implies  that CP violation can be very large
in  these  channels.  Instead,  in  $b  {\bar  b}$ fusion,  the  ratio
$\hat{\sigma}_2/\hat{\sigma}_1$   is   always   less  than   1\%,   so
CP-violating  effects in  the production  of  longitudinally polarized
$\tau$ leptons are unobservably small in this case.

The smallness  of $\hat{\sigma}_2$ in $b {\bar  b}$ fusion is a result
of    an    intriguing   interplay   between    unitarity     and  CPT
invariance~\cite{APNPB}.   In  detail,  the CP-violating cross section
$\hat{\sigma}_2$ may be calculated by
\begin{equation}
\hat{\sigma}_2 (b\bar{b}\to H\to \tau^+\tau^-) \ =\ \frac{1}{4}\,
\bigg[\, \hat{\sigma} (b\bar{b} \to H \to \tau^+_R \tau^-_R) \ -\
\hat{\sigma} (b\bar{b} \to H \to \tau^+_L\tau^-_L)\, \bigg]\,,
\end{equation}
where $\hat{\sigma}$ denotes  the usual subprocess cross section.  For
the scenario under study, unitarity cuts  of $b\bar{b}$ pairs dominate
the  absorptive part of  the Higgs-boson self-energies. Employing this
fact and the optical theorem, we obtain
\begin{eqnarray}
  \label{s2CP}
\sum\limits_{\lambda = L,R}\ \hat{\sigma}_2 (b_\lambda \bar{b}_\lambda
\to H \to \tau^+\tau^-) &=&  C_{\rm PS}\, \bigg( \, \imag\, {\cal
T} (\tau^+_R \tau^-_R \to H \to \tau^+_R \tau^-_R) \nonumber\\ 
&& \hspace{-3cm}-\ \imag\, {\cal T} (\tau^+_L \tau^-_L \to H \to \tau^+_L
\tau^-_L)\,\bigg)\quad +\quad {\cal O}\Big[\hat{\sigma}'_2\,B(H_{1,2,3}\to
\tau^+\tau^-)\Big]\, , \qquad
\end{eqnarray}
where  $C_{\rm PS}$  is  a phase-space  correction  factor and  ${\cal
T}(\tau^+_{L,R}  \tau^-_{L,R} \to  H  \to \tau^+_{L,R}  \tau^-_{L,R})$
denote the usual  matrix elements.  In~(\ref{s2CP}), $\hat{\sigma}'_2$
is  the  CP-violating  cross-section  $\hat{\sigma}_2$  calculated  by
omitting  the   off-diagonal  absorptive  parts   in  the  Higgs-boson
propagator  matrix  $D(\hat{s})$.  The  size of  $\hat{\sigma}'_2$  is
smaller at least  by a factor 10 than  the spin-averaged cross-section
$\hat{\sigma}_1$.   On  the other  hand,  CPT  invariance imposes  the
constraint
\begin{equation}
  \label{CPTconstraint}
{\cal T} (\tau^+_R \tau^-_R \to H \to \tau^+_R \tau^-_R)\ =\ 
{\cal T} (\tau^+_L \tau^-_L \to H \to \tau^+_L \tau^-_L)\, .
\end{equation}
With the aid  of  (\ref{CPTconstraint}), it is    not   difficult to see
using~(\ref{s2CP}) that the CP-violating cross section $\hat{\sigma}_2
(b\bar{b}\to H\to  \tau^+\tau^-)$ vanishes   up to  CP-violating terms
suppressed by extra factors of order $B(H_{1,2,3}\to \tau^+\tau^-)$.

Our numerical estimates presented  in Fig.~\ref{fig:3mix34}  show that
the     CP-violating       transverse-polarization     cross   section
$\hat{\sigma}_{4}$ may be  quite sizeable for all production channels.
However, $\hat{\sigma}_{4}$ generically  exhibits an alternating  sign
for  $b\bar{b}$ and  $gg$  collisions, and CP   violation becomes very
small  when we integrate over the  whole Higgs-boson resonance region.
Moreover, the transverse $\tau^\pm$ polarizations will be difficult to
measure at the LHC because of the experimental conditions, notably the
large boosts     of the $\tau^\pm$.   On   the  other  hand, analogous
asymmetries might  be observable in  ${\bar t} t$ production and/or in
$\tau^+ \tau^-$ production at a $\mu^+ \mu^-$ collider.

The Higgs production channels via $b\bar{b}$ and $gg$ fusion processes
can be separated from the $W^+W^-$ fusion channel by applying a number
of kinematic  cuts~\cite{ATLAS} including the imposition of  a veto on
any  hadronic  activity  between jets~\cite{JF0,JF1}.   Therefore,  we
treat  the contributions from  $b\bar{b}$ and  $gg$ collisions  to the
physical  Higgs-exchange  process  $pp   \to  H  \to  \tau^+\tau^-  X$
separately from  those coming from $WW$ fusion.   More explicitly, the
physical $\tau^+ \tau^-$ cross  section can be computed by integrating
the parton-level  cross sections with the distribution  of $b$ quarks,
gluons and $W$-bosons in the proton,
\begin{eqnarray}
  \label{xggbb}
\tau\ \frac{d\sigma_{\rm tot}}{d\tau}\, \Big( pp\, (b\bar{b},gg)\ \to\ 
\tau^+\tau^- X \Big)  &=&
 4\, \hat{\sigma}_1(b\bar{b}\rightarrow H \rightarrow \tau^+\tau^-) \
 \tau\,\frac{d{\cal L}^{bb}}{d\tau}\nonumber\\
&&+\, 4\ K\,\hat{\sigma}_1(gg\rightarrow H \rightarrow \tau^+\tau^-) \
\tau\,\frac{d{\cal L}^{gg}}{d\tau}\,,\\[3mm]
  \label{xWW}
\tau\,\frac{d\sigma_{\rm tot}}{d\tau}\,\Big( pp\, (W^+ W^-)\ \to\
\tau^+\tau^- X\Big) &=& 
\, 4\, \hat{\sigma}_1( W^+ W^-
\rightarrow H \rightarrow \tau^+\tau^-)\ \tau\,\frac{d{\cal
L}^{WW}}{d\tau}\, , 
\end{eqnarray}
where $\tau$  is the Drell--Yan  variable $\tau=\hat{s}/s$ and  $s$ is
the  invariant   squared  centre-of-mass   energy  of  the   LHC.   In
(\ref{xggbb}),  we use the  value $K=1+\frac{\alpha_s(\hat{s})}{\pi}\,
(\pi^2+11/2)$, ignoring  the small difference  between the $K-$factors
for CP-even  and CP-odd Higgs states.  The  effective luminosities for
$b\bar{b}$ and  $gg$ collisions, ${\cal L}^{bb}$  and ${\cal L}^{gg}$,
may be determined by
\begin{eqnarray}
 \tau\frac{d{\cal L}^{bb}}{d\tau} &=&
 \int_{\tau}^1 dx\, \left[\,\frac{\tau}{x}\,
        b(x,Q) \,
        \bar{b}\left(\frac{\tau}{x},Q\!\right)\
        +\ (b\leftrightarrow \bar{b})\, \right]  \,,\nonumber \\
\tau\frac{d{\cal L}^{gg}}{d\tau} &=&
\int_{\tau}^1dx\ \frac{\tau}{x}\,g(x,Q)\,g\left(\frac{\tau}{x},Q\right)\,,
\end{eqnarray}
where $b(x,Q)$, $\bar{b}(x,Q)$   and $g(x,Q)$ are  the $b$, $\bar{b}$
and gluon   distribution functions  in   the proton  and $Q$   is  the
factorization   scale.   In  our    numerical  analysis,  we   use the
leading-order CTEQ6L~\cite{CTEQ6}  parton   distribution functions for
$b(x,Q)$ and $\bar{b}(x,Q)$,    and  the CTEQ6M  parton   distribution
function  for   $g(x,Q)$.     We   choose  the   factorization   scale
$Q=\sqrt{\hat{s}}/4$ for the   $b$-quark  fusion  process as  suggested   and
confirmed in~\cite{BBH2}.

Correspondingly,  in (\ref{xWW}), the  effective luminosities  for the
transverse and longitudinal  $W$-bosons, denoted as $W^\pm_{T,L}$, can
be computed in terms of effective densities $F_{W_{T,L}^\pm}^p\!(x,Q)$
in the colliding protons, which are in turn calculated in terms of the
quark    parton    distribution     functions    $q(x,Q)$    in    the
proton:\footnote{Here  we  consider identical  polarizations  for  the
$W^\pm$ bosons in the $W^+W^-$ fusion process.}
\begin{eqnarray}
\tau\ \frac{d{\cal L}^{W_P W_P}}{d\tau} &=&
\int_\tau^1 dx\ 
 \left[\,\frac{\tau}{x}\,
        F_{W_P^+}^p\!(x,Q) \,
        F_{W_P^-}^p\!\!\left(\frac{\tau}{x},Q\!\right)\
        +\ (W_P ^+\leftrightarrow W_P^-)\, \right]\,,\nonumber\\
F_{W_P^+}^p(x,Q) &=& \sum_{q=u,\bar{d},c,\bar{s}}\
\int_x^1 \frac{dy}{y} \ q(y,Q)\,F^q_{W^+_P}(x/y,Q)\, ,
\end{eqnarray}
where  the  transverse  $(P=T)$  and  longitudinal  $(P=L)$  effective
densities   $F^q_{W^+_{T,L}}$  in   the   quark  $q$   are  given   by
\cite{WWF1,WWF2}
\begin{eqnarray}
F^q_{W^+_T}(x,Q) &=&\frac{\alpha_{\rm em}}{8\pi\,s_W^2}\
\ln\left(\frac{Q^2}{M_W^2}\right)\
\frac{1+(1-x)^2}{x}\ ,\nonumber\\ 
F^q_{W^+_L}(x,Q) &=& \frac{\alpha_{\rm em}}{4\pi\,s_W^2}\ \frac{1-x}{x}\ .
\end{eqnarray}
Note that the summation over  quark flavours $q$ in the expression for
$F_{W_{T,L}^-}^p(x,Q)$ includes  $q=\bar{u},d,\bar{c},s$. Moreover, we
take $Q = \sqrt{\hat{s}}$ in our numerical estimates.

To   analyze  the signatures  of  CP  violation  in  the production of
longitudinally polarized  $\tau$-leptons, we first define the physical
observables
\begin{equation}
\sigma_{RR}\ =\ \sigma(pp\ \to\ H\ \to\  \tau^+_R\tau^-_R X) \,, \qquad
\sigma_{LL}\ =\ \sigma(pp\ \to\ H\ \to\  \tau^+_L\tau^-_L X) \, .
\end{equation}
Evidently, the total cross section for Higgs production and decay
into  $\tau^+\tau^-$  pairs is  given in   terms of $\sigma_{RR}$  and
$\sigma_{LL}$ by
\begin{equation}
\sigma_{\rm tot} (pp\ \to\  H\ \to\ \tau^+\tau^- X)\ = \ \sigma_{RR}\ +\
\sigma_{LL}\ .
\end{equation}
Although the initial  state $pp$ is  not  symmetric under CP,  it can,
however, be shown  that,  up to negligible  higher-order  CP-violating
electroweak effects,  the effective luminosities  for $gg$, $b\bar{b}$
and $W^+W^-$  densities will  be practically the   same  for $pp$  and
$\bar{p}\bar{p}$  collisions.   Therefore,  the  difference of   cross
sections
\begin{equation}
  \label{DeltaCP} 
\Delta\sigma_{\rm CP}\ =\ \sigma_{RR}\ -\ \sigma_{LL}
\end{equation}
is  a  measure   of genuine  CP  violation at   the   LHC.  In analogy
with~(\ref{xggbb}) and~(\ref{xWW}),  the CP-violating  cross   section
$\Delta\sigma_{\rm CP}$ can be computed by
\begin{eqnarray}
  \label{xCPggbb}
\tau\ \frac{d\Delta\sigma_{\rm CP}}{d\tau}\, \Big( pp\, (b\bar{b},gg)\ \to\ 
\tau^+\tau^- X \Big)  &=&
 4\,\hat{\sigma}_2\, (b\bar{b}\rightarrow H \rightarrow \tau^+\tau^-) \
 \tau\,\frac{d{\cal L}^{bb}}{d\tau}\nonumber\\
&&+\ 4\, K\,\hat{\sigma}_2\, (gg\rightarrow H \rightarrow \tau^+\tau^-) \
\tau\,\frac{d{\cal L}^{gg}}{d\tau}\,,\\[3mm]
  \label{xCPWW}
\tau\,\frac{d\Delta\sigma_{\rm CP}}{d\tau}\,\Big( pp\, (W^+ W^-)\ \to\
\tau^+\tau^- X\Big) &=& 4\, \hat{\sigma}_2 ( W^+ W^- \to H \to \tau^+\tau^-)\ 
\tau\,\frac{d{\cal L}^{WW}}{d\tau}\, .\qquad
\end{eqnarray}
To  gauge the sizes of  the signatures of CP violation   at the LHC, we
define the following two CP asymmetries:
\begin{equation}
  \label{CPasym}
a_{\rm CP} (\tau)\ \equiv \ \frac{\tau\ 
\frac{\displaystyle d\Delta\sigma_{\rm CP}}{\displaystyle d\tau}}{
\tau\ \frac{\displaystyle d\sigma_{\rm tot}}{\displaystyle d\tau}}\ ,
\qquad\qquad
{\cal A}_{\rm CP} \ \equiv \ 
\frac{ \Delta\sigma_{\rm CP} }{\sigma_{\rm tot}}\ ,
\end{equation}
pertinent to the hadron-level  processes $pp\, (b\bar{b},gg,WW)\  \to\
H\ \to\ \tau^+\tau^- X$.

\begin{figure}[p]
\epsfig{figure=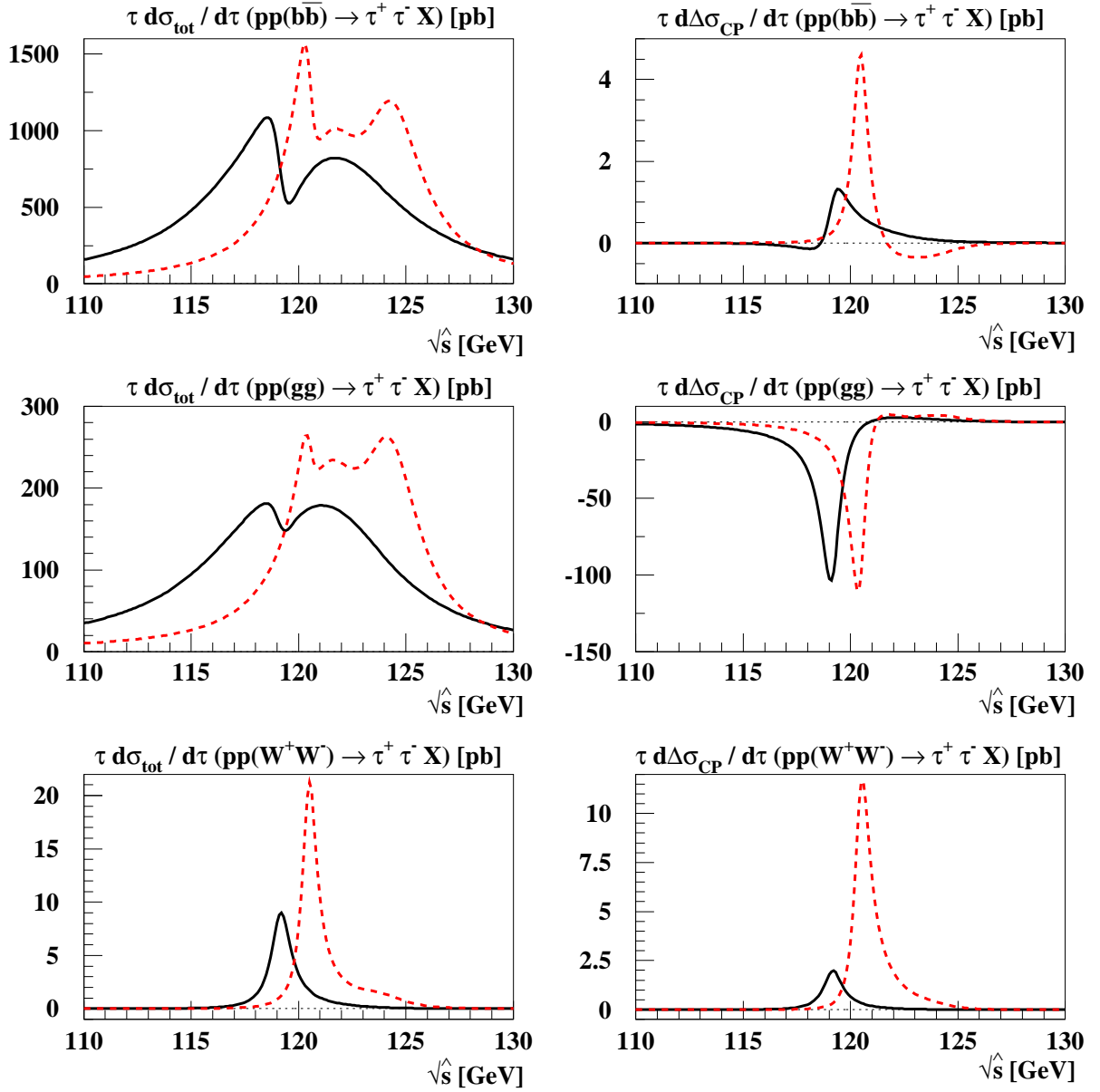,height=18cm,width=18cm}
\vspace{-0.7cm}
\caption{\it The differential cross sections $\tau\,
\frac{d\sigma_{\rm tot}}{d\tau}$ and $\tau\, \frac{d\Delta\sigma_{\rm
CP}}{d\tau}$ as functions of $\sqrt{{\hat s}}$. The upper frames are
for the process $b\bar{b}\rightarrow H \rightarrow \tau^+\tau^-$, the
middle ones for $gg\rightarrow H \rightarrow \tau^+\tau^-$ and the
lower ones for $W^+W^-\rightarrow H \rightarrow \tau^+\tau^-$.  The
solid lines are for the three-Higgs mixing scenario with
$\Phi_3=-90^\circ$ and the dashed ones with 
$\Phi_3=-10^\circ$.}\label{fig:cxtd}
\end{figure}

\begin{figure}[p]
\epsfig{figure=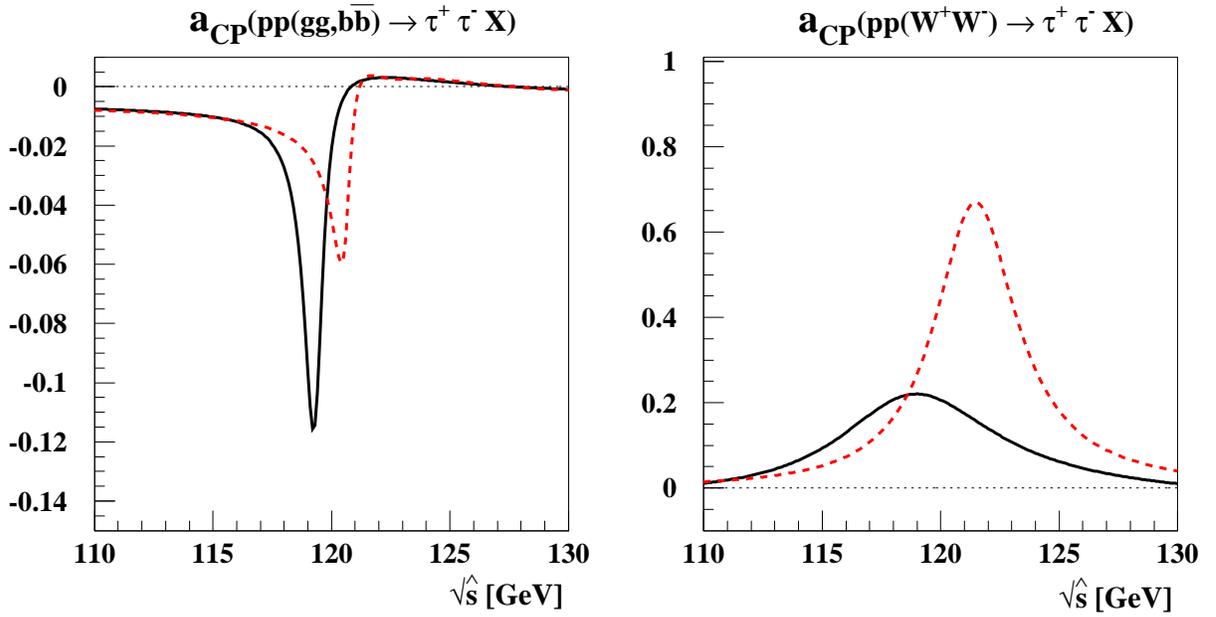,height=18cm,width=18cm}
\vspace{-8cm}
\caption{\it Numerical  estimates   of differential   CP   asymmetries
$a_{\rm  CP}$  defined in~(\ref{CPasym})  as functions of $\sqrt{{\hat
s}}$.  The solid  line corresponds to the  three-Higgs mixing scenario   with
$\Phi_3=-90^\circ$ and the dashed one to $\Phi_3=-10^\circ$.  }
\label{fig:ratio}
\end{figure}

We plot in Fig.~\ref{fig:cxtd} the differential cross sections $\tau\,
\frac{d\sigma_{\rm tot}}{d\tau}$  and $\tau\, \frac{d\Delta\sigma_{\rm
CP}}{d\tau}$ as  functions of $\sqrt{\hat{s}}$.  The  upper two frames
are for the process  $b\bar{b}\rightarrow \tau^+\tau^-$, the frames in
the middle  for $gg\rightarrow \tau^+\tau^-$,  and the lower  ones for
$W^+W^- \to  \tau^+\tau^-$.  We observe  that the the  main production
mechanism is  $b {\bar b}$ fusion,  which gives a  cross section about
five times larger than that due to gluon fusion for the scenario under
consideration.    However,   as   has   been  mentioned   above,   the
$W^+W^-$-fusion  production  mechanism, albeit  much  smaller, can  be
experimentally  distinguished from  that  due to  $b\bar{b}$ and  $gg$
collisions.   Therefore,  in Fig.~\ref{fig:ratio}  we  display the  CP
asymmetry $a_{\rm CP}$ defined in (\ref{CPasym}) separately for the $b
{\bar b} + gg$ and $W^+  W^-$ subprocesses.  We note that the large CP
asymmetry in $gg$ subprocess is  diluted by the dominant cross section
via $b\bar{b}$ fusion\footnote{Specifically, the total CP asymmetry in
the $gg$ subprocess  is ${\cal A}_{\rm CP}^{gg}=-8.4\,(-6.2)\,\%$, for
$\Phi_3=-90^\circ\,(-10^\circ)$.   However,  after  the  inclusion  of
$b\bar{b}$  collisions,  the  combined  CP  asymmetry  ${\cal  A}_{\rm
CP}^{b\bar{b}+gg}$ reduces to $-1.4\,(-1.0)\,\%$ .}.

It  is important to   emphasize here that the CP-violating  observable
$\tau\, \frac{d\Delta\sigma_{\rm  CP}}{d\tau}$ for the 
$WW$-fusion process does not change sign as
the   $\tau^+\tau^-$-system  energy $\sqrt{\hat{s}}$  varies over  the
entire  Higgs-boson resonant    region.  
Such a kinematic   behaviour is ensured by the presence of
the off-diagonal absorptive parts of the Higgs-boson
self-energies.
Instead, if these off-diagonal absorptive
parts    are neglected, we     find  the erroneous   result   that the
CP-violating observable  flips sign  in the  resonant  region, thereby
leading  to unobservably small CP asymmetries   when averaged over the
energy $\sqrt{\hat{s}}$.

Although CP violation in the  $WW$ and $gg$ production channels may be
sizeable,  it is difficult  to measure  the differential  CP asymmetry
$a_{\rm CP}$  at the LHC because  of the low energy  resolution of the
reconstructed  $\tau^+\tau^-$  invariant mass.   This  last fact  also
limits our  ability to reconstruct  with sufficient accuracy  the line
shape of the decaying coupled  Higgs-boson system at the LHC.  This is
unfortunate since one would miss the very interesting feature shown in
Fig.~\ref{fig:cxtd} that,  unlike the case of a  single resonance, the
locations of the various maxima  in the resonant line shapes described
by  $\tau\, \frac{d\sigma_{\rm tot}}{d\tau}$  crucially depend  on the
production   and   decay   channels   of   the   coupled   Higgs-boson
system.  Therefore,   the  extra  analyzing  power   of  $e^+e^-$  and
$\mu^+\mu^-$ colliders  would be  highly valuable for  unravelling the
existence of a strongly-mixed  Higgs-boson system and studying in more
detail its dynamical properties.

\begin{figure}[p]
\epsfig{figure=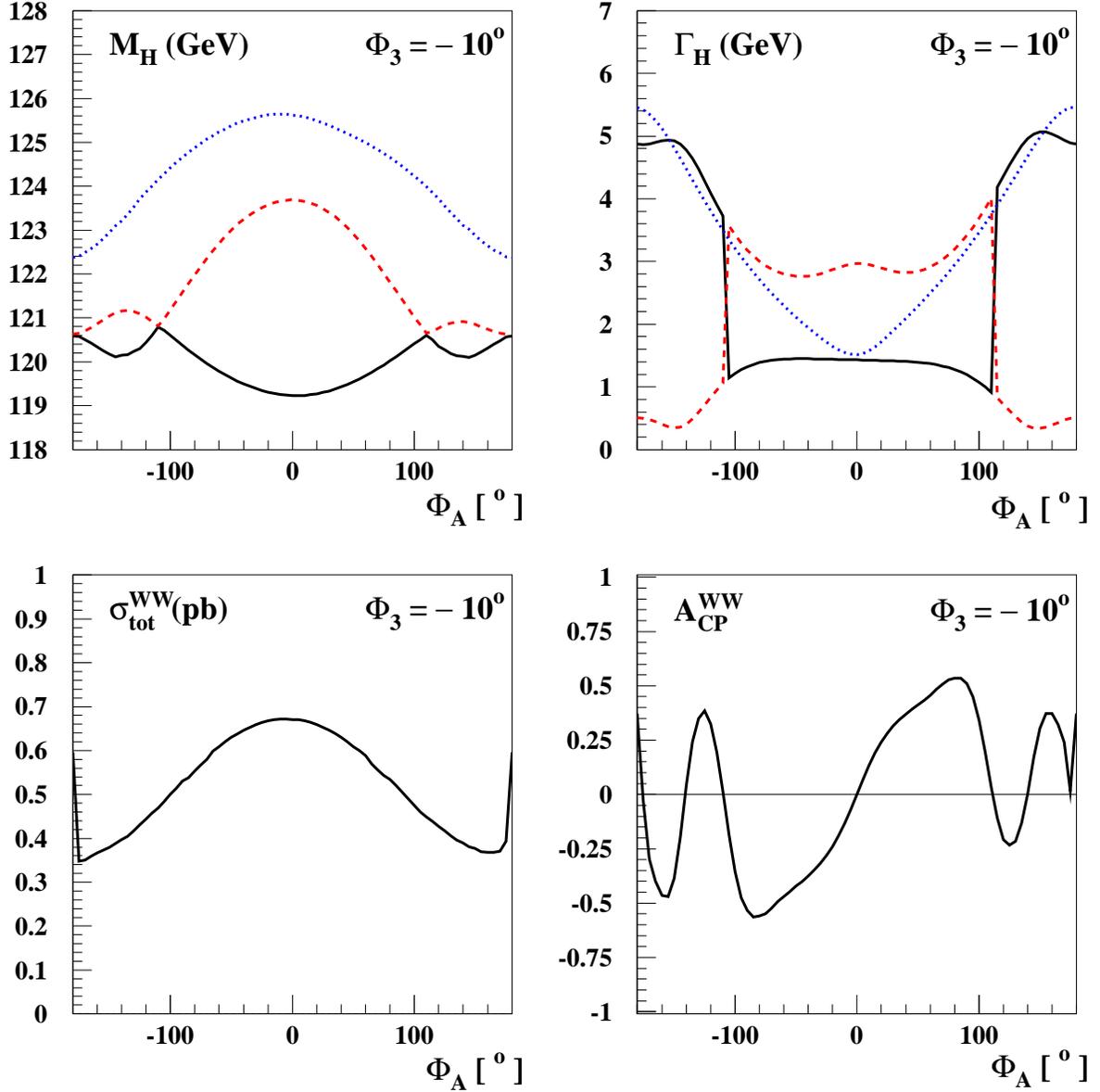,height=18cm,width=18cm}
\caption{\it  Numerical  estimates  of  Higgs-boson masses  and  decay
widths,  the  cross-section  $\sigma_{\rm   tot}  (pp(WW)  \to  H  \to
\tau^+\tau^-  X)$  and  its   associated  total  CP  asymmetry  ${\cal
A}^{WW}_{\rm CP}$ defined in~(\ref{CPasym})  as functions of $\Phi_A =
\Phi_{A_t} =  \Phi_{A_b} =  \Phi_{A_\tau}$, for $\Phi_3  = -10^\circ$.
In the upper  two frames, the solid, dashed and  dotted  lines 
refer to the
$H_1$, $H_2$ and $H_3$ bosons, respectively.}\label{fig:p3m10}
\end{figure}

\begin{figure}[p]
\centerline{\epsfig{figure=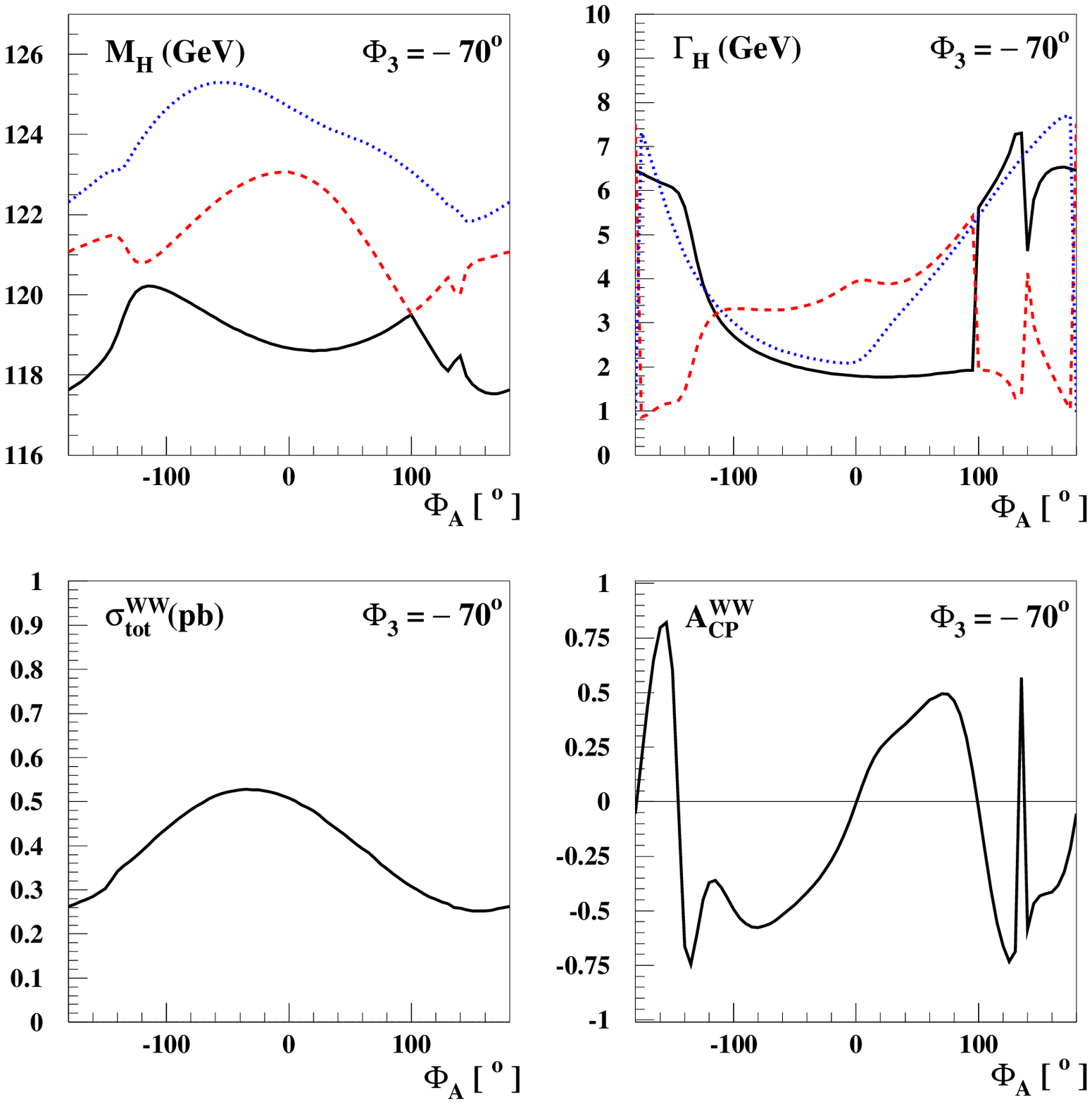,height=18cm,width=18cm}}
\caption{\it  The same as in Fig.~\ref{fig:p3m10}, but for
$\Phi_3 = -70^\circ$.}\label{fig:p3m70}
\end{figure}

\begin{figure}[p]
\epsfig{figure=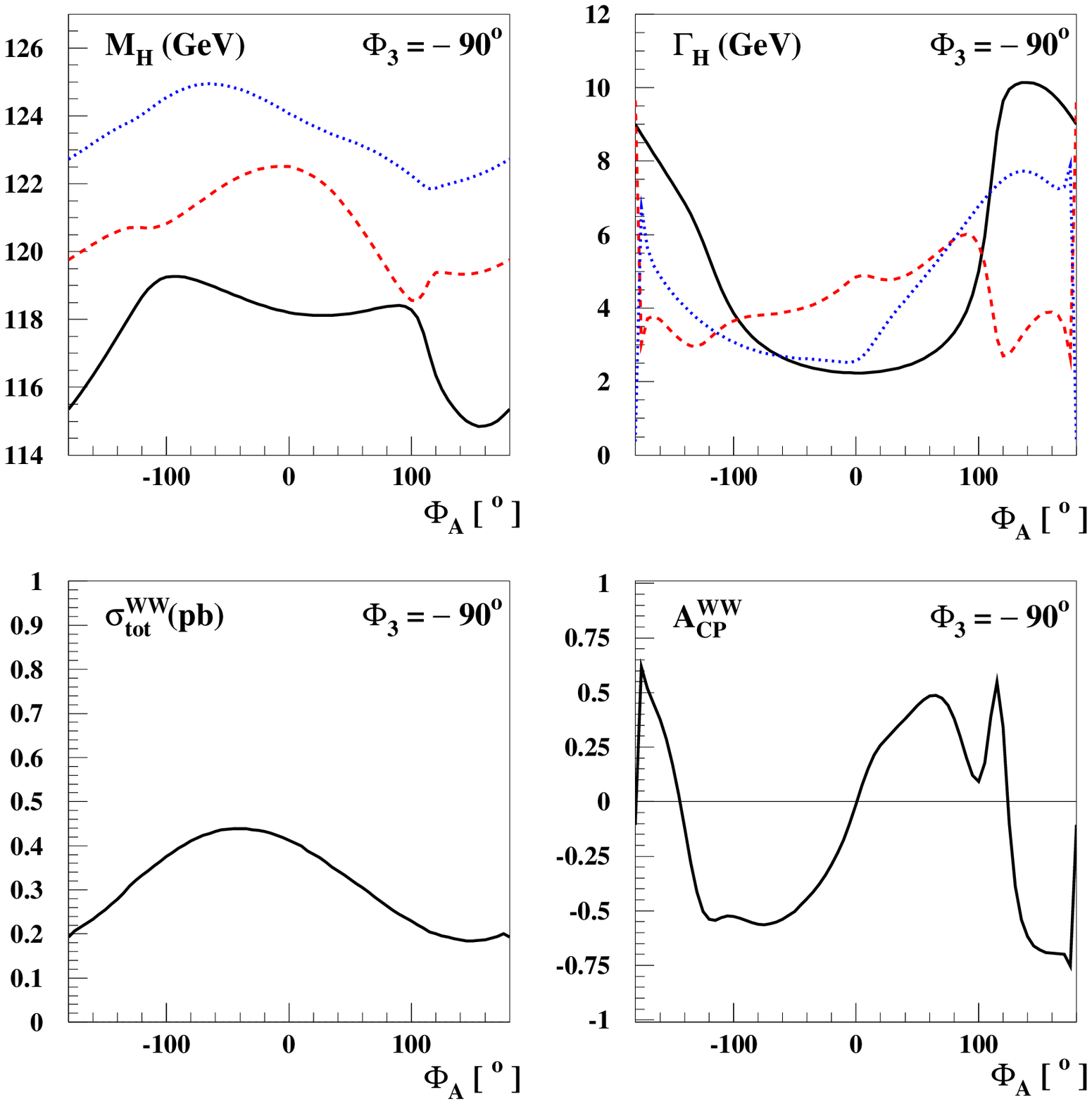,height=18cm,width=18cm}
\caption{\it  The same as in Fig.~\ref{fig:p3m10}, but for
$\Phi_3 = -90^\circ$.}\label{fig:p3m90}
\end{figure}

\begin{figure}[p]
\epsfig{figure=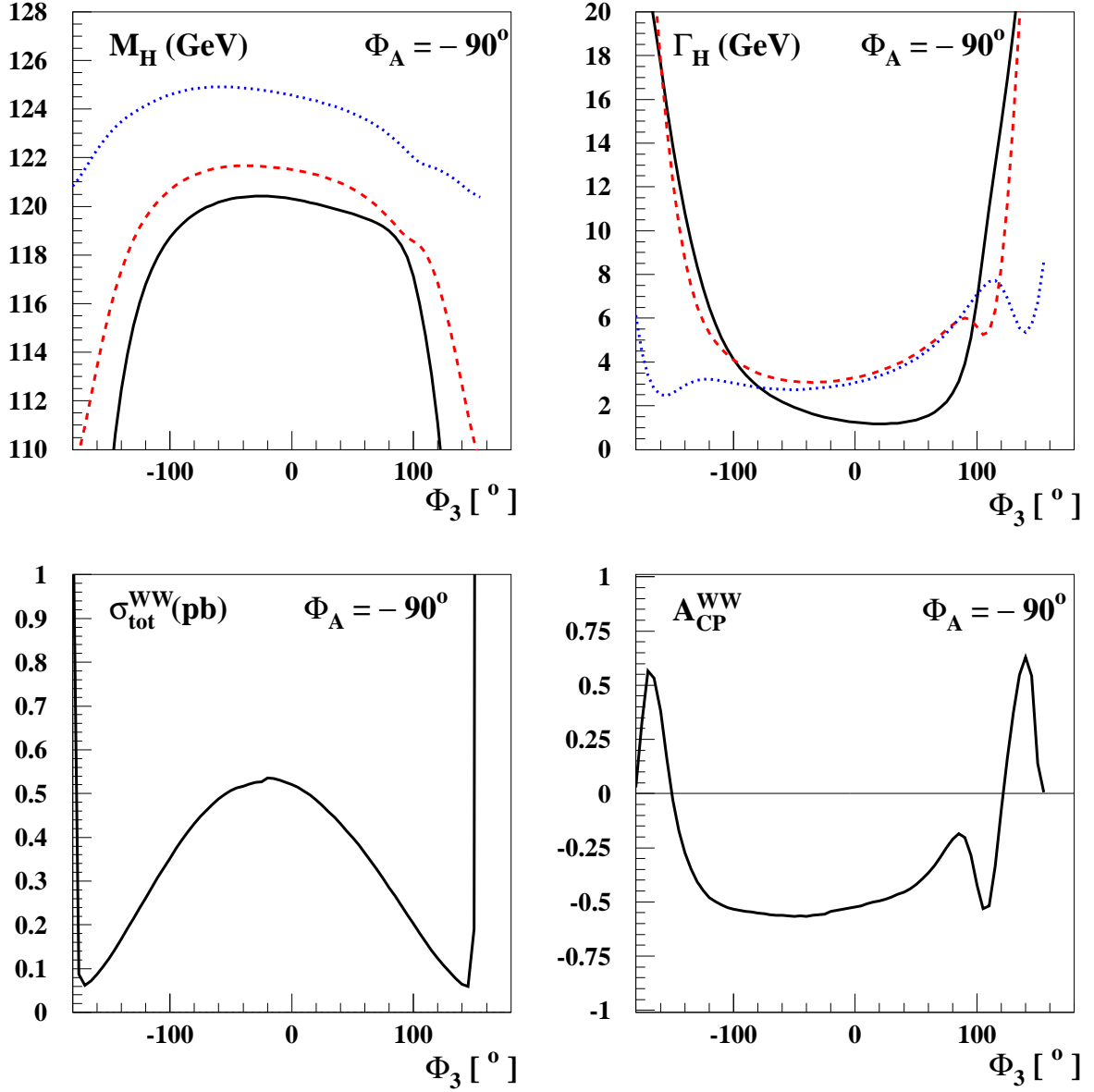,height=18cm,width=18cm}
\caption{\it     Numerical    values    for     $M_{H_{1,2,3}}$    and
$\Gamma_{H_{1,2,3}}$, $\sigma_{\rm tot} (pp(WW) \to H \to \tau^+\tau^-
X)$ and ${\cal A}^{WW}_{\rm CP}$ as functions of $\Phi_3$, for $\Phi_A
= \Phi_{A_t} = \Phi_{A_b} = \Phi_{A_\tau} = -90^\circ$.  We follow the
same line conventions as in Fig.~\ref{fig:p3m10}.}\label{fig:pam90}
\end{figure}

Motivated   by   the   large   differential  CP   asymmetry   in   the
$W^+W^-$-fusion process, we perform  a numerical analysis of the total
CP  asymmetry  for  the  reaction  $pp(WW) \to  H  \to  \tau^+\tau^-$,
integrated   over  the   Higgs   resonance  peaks.    We  present   in
Figs.~\ref{fig:p3m10},~\ref{fig:p3m70}     and~\ref{fig:p3m90}     the
predicted values for the cross-section $\sigma_{\rm tot} (pp(WW) \to H
\to \tau^+\tau^-  X)$ and its associated total  integrated CP asymmetry
${\cal  A}^{WW}_{\rm CP}$  defined in~(\ref{CPasym})  as  functions of
$\Phi_A  = \Phi_{A_t}  = \Phi_{A_b}  = \Phi_{A_\tau}$,  for  $\Phi_3 =
-10^\circ$, $-70^\circ$, and  $-90^\circ$, respectively.  In the upper
two  frames  of  the  figures,   we  display  the  dependence  of  the
Higgs-boson masses  and their decay  widths on the  CP-violating phase
$\Phi_A$, where the solid, dashed and dotted lines refer to the $H_1$,
$H_2$ and  $H_3$ bosons, respectively.  In our  numerical analysis, we
fix the remaining parameters  of the MSSM as in~(\ref{MSSM1}).  Unlike
in   Figs.~\ref{fig:p3m10},~\ref{fig:p3m70},  and~\ref{fig:p3m90},  we
present  in  Fig.~\ref{fig:pam90} numerical  estimates  by fixing  the
value of  $\Phi_A$ to $-90^\circ$, but varying  the CP-violating phase
$\Phi_3$.  For  the scenario under  study, all three Higgs  bosons mix
among themselves significantly, giving  rise to level crossings as the
CP-odd  phases  vary.  These  effects  of  level  crossing lead  to  a
non-trivial behaviour  in $\Gamma_{H_i}$, which  is between 1  GeV and
10~GeV~\footnote{In Fig.~\ref{fig:pam90}, the  widths of the $H_1$ and
$H_2$ become larger than 10~GeV when $\Phi_3 > 100^\circ$ or $\Phi_3 <
-140^\circ$,  where   $M_{H_1}$  decreases  very   rapidly  and  $H_1$
decouples from the $H_2 -  H_3$ mixing system.}, and in ${\cal A}_{\rm
CP}^{WW}$.  We find that the total cross section is between 0.1 pb and
0.7 pb and is comparable to  the corresponding SM cross section 0.3 pb
for  $M_{H_{\rm SM}}=120$~GeV~\cite{ATLAS}.   We observe  that  the CP
asymmetry ${\cal  A}^{WW}_{\rm CP}$  is large for  a wide range  of CP
phases and can even be  as large as 80\% for $\Phi_3=-70^\circ$.  Even
for  small CP-violating phases,  $\Phi_3=-10^\circ$ and  $(180^\circ -
|\Phi_A|) < 20^\circ$,  the CP asymmetry can be  $\sim$~50\%, as shown
in Fig.~\ref{fig:p3m10}.   Again, we note  that possible uncertainties
in  the calculation of  the cross  sections largely  cancel in  the CP
asymmetry ${\cal A}_{\rm CP}$.

Finally,  we comment  briefly  on the  possible  impact of  low-energy
constraints on  the CP asymmetries, especially those  arising from the
non-observation of  the electron and  neutron EDMs and the  absence of
the  Higgs-mediated  $B$-meson  decay  $B_{s,d}  \to  \mu\mu$  at  the
Tevatron~\cite{CDFmumu}.   The  EDM  constraints may  be  considerably
relaxed if  we consider  scenarios with the  first two  generations of
squarks  heavier than  about 3~TeV,  and if  we allow  some  degree of
cancellations~\cite{EDM2}  between   the  one-  and   higher-loop  EDM
contributions~\cite{EDMnote}.  For the  scenarios under study, we have
estimated that  the required degree of cancellation  is always smaller
than     80\%,     where     100\%     corresponds     to     complete
cancellation. Therefore, a full implementation of EDM constraints will
not alter the results of the present analysis in a significant way. On
the other hand, the lack of observation of $B_{s,d} \to \mu\mu$ at the
Tevatron~\cite{CDFmumu} imposes further  constraints on the parameters
of the CP-violating MSSM.  However, the derived constraints are highly
flavour-dependent and can be  dramatically relaxed for certain choices
of  the  soft  SUSY-breaking   mass  spectrum  that  enable  unitarity
cancellations   in  the   flavour  space.    For  a   detailed  study,
see~\cite{DP}.


\section{Conclusions and Prospects}\label{sec:conclusions}

We have presented  the general  formalism for analyzing   CP-violating
phenomena in the  production, mixing and decay  of a coupled system of
multiple CP-violating  neutral Higgs bosons.   Our formalism, which is
developed from~\cite{APNPB}, can be applied to models with an extended
CP-violating Higgs sector,  including the highly predictive  framework
of  the MSSM with radiative Higgs-sector  CP  violation.  An important
element  of   the   formalism  is the   consideration    of  the  full
$s$-dependent $3\times   3$ Higgs-boson  propagator matrix,  where the
gauge-mediated contributions  to self-energies have been calculated in
the framework of the Pinch Technique~\cite{PTgeneral,PINCH}.

As  an application  of our formalism,  we have  studied  in detail the
production of CP-violating MSSM $H_{1,2,3}$  bosons via ${\bar b}  b$,
$gg$ and $W^+ W^-$ collisions and their subsequent decays into $\tau^+
\tau^-$  pairs  at  the  LHC.  In addition  to   the Higgs self-energy
effects, we have  also given explicitly the relevant formulae in  the
MSSM with   loop-induced  CP violation  in   the production  and decay
vertices.  We  have considered specific   MSSM scenarios that  predict
three  nearly  degenerate,    strongly-mixed    Higgs  bosons     with
$M_{H_{1,2,3}} \sim 120$~GeV.   Such scenarios  naturally occur in   a
general CP-violating  MSSM when $\tan\beta$ is  larger than 30 and the
charged  Higgs boson  is lighter  than  about 160~GeV.  

We    have analyzed  CP  asymmetries    in  both  longitudinally-  and
transversely-polarized $\tau^+ \tau^-$ pairs. CP asymmetries that make
use of  the  transverse polarization  of   the $\tau$-lepton, although
being  intrinsically very large   in  the CP-violating MSSM  scenarios
mentioned above, generically  exhibit an  alternating sign and  become
unobservably small after     averaging over  the entire    Higgs-boson
resonant  region.   Also,   reconstruction of  transversely  polarized
$\tau$ leptons appears rather difficult  at the LHC.  However, such CP
asymmetries might ideally be  tested at a $\mu^+\mu^-$ collider, where
a high energy resolution can be achieved.

At  the  LHC,   more  promising  are  CP  asymmetries   based  on  the
longitudinal  $\tau$-lepton  polarization.    In  particular,  the  CP
asymmetry in  the production channel $W^+W^- \to  H_{1,2,3} \to \tau^+
\tau^-$ may  well exceed the 10\%  level and reach values  up to 80\%.
It is important  to stress again that the  $WW$ production channel can
be cleanly isolated  from the $gg$ and $b\bar{b}$  channels, mainly by
vetoing   any   hadronic   activity   between   jets   (for   details,
see~\cite{ATLAS}).  Hence, depending on the efficiency of longitudinal
$\tau$-lepton   polarization  techniques~\cite{Was},   the  production
channel  $W^+W^-  \to H_{1,2,3}  \to  \tau^+  \tau^-$  may become  the
`golden' channel for studying  signatures of Higgs-sector CP violation
at the~LHC.

The  formalism presented in this paper  may easily be applied to other
colliders as  well,  most  notably to  $e^+e^-$,   $\gamma \gamma$ and
$\mu^+ \mu^-$ colliders.    At $e^+e^-$ linear  colliders, Higgs
bosons can  copiously be produced via  the  Higgsstrahlung or $W^+W^-$
fusion processes.  At $\gamma\gamma$  and $\mu^+\mu^-$ colliders,  the
polarizations  of  the colliding  beams  may  also  be varied, thereby
providing   additional probes  of    Higgs-sector CP  violation.   The
aforementioned colliders can  provide cleaner experimental conditions
than the LHC.  Consequently, even if the CP asymmetries discussed here
prove difficult to observe at the LHC,  the formalism and the analysis
techniques developed here   to investigate Higgs-sector   CP violation
will be directly applicable to such future colliders as well.

\subsection*{Acknowledgements}
We  thank Jeff Forshaw  for  discussions.  The work of   JSL and AP  is
supported in part by the PPARC research grant PPA/G/O/2000/00461.

\end{document}